\documentclass[pageno]{sig-alternate}

\usepackage{mathptmx} 

\newcommand{\ignore}[1]{}
\usepackage{fancyhdr}
\usepackage[normalem]{ulem}
\usepackage[hyphens]{url}
\usepackage{microtype}
\usepackage{gensymb}
\usepackage[colorlinks = true,
            linkcolor = blue,
            urlcolor  = blue,
            citecolor = blue,
            anchorcolor = blue]{hyperref}
\usepackage{subfigure}

\newcommand{\footnoteremember}[2]{
\footnote{#2}
\newcounter{#1}
\setcounter{#1}{\value{footnote}}
}

\fancypagestyle{firstpage}{
  \fancyhf{}
  
  \fancyfoot[C]{\thepage}
}

\title{\vspace{-3.5ex}Mixed-Signal Charge-Domain Acceleration of Deep Neural Networks through Interleaved Bit-Partitioned Arithmetic\vspace{-5ex}} 

\vspace{5ex}
\author{
	\vspace{0.05in}
	\fontsize{10}{12}\selectfont{}Soroush Ghodrati \quad \fontsize{10}{12}\selectfont{}Hardik Sharma$^\dag$ \quad \fontsize{10}{12}\selectfont{}Sean Kinzer \quad \fontsize{10}{12}\selectfont{}Amir Yazdanbakhsh$^\ddag$\thanks{\enskip This work has been done when the author was a PhD student at Georgia Institute of Technology.} \\ 
	\vspace{0.05in}
	 \quad \fontsize{10}{12}\selectfont{}Kambiz Samadi$^\flat$ \quad \enspace \fontsize{10}{12}\selectfont{}Nam Sung Kim$^\P$ \quad \fontsize{10}{12}\selectfont{}Doug Burger$^\S$ \quad \fontsize{10}{12}\selectfont{}Hadi Esmaeilzadeh\\
	\vspace{0.05in}
	\fontsize{10}{12}\selectfont{}\textbf{A}lternative \textbf{C}omputing \textbf{T}echnologies ({\color[HTML]{0B6121}{\textbf{ACT}}}) Lab \\
	\vspace{0.05in}
	\fontsize{10}{12}\selectfont{}University of California, San Diego  \\ 	\vspace{0.05in}
	 \fontsize{10}{12}\selectfont{}$^\dag$Georgia Institute of Technology \quad \fontsize{10}{12}\selectfont{}$^\ddag$Google Research  
	\quad \fontsize{10}{12}\selectfont{}$^\flat$Qualcomm Technologies \quad    \fontsize{10}{12}\selectfont{}$^\P$Samsung Electronics \quad \fontsize{10}{12}\selectfont{}$^\S$Microsoft \\
	\vspace{0.05in} 
\textcolor{blue}{\textsf{
\href{mailto:soghodra@eng.ucsd.edu}{soghodra@eng.ucsd.edu} \quad  \href{mailto:hsharma@gatech.edu}{hsharma@gatech.edu} \quad \href{mailto:skinzer@eng.ucsd.edu}{skinzer@eng.ucsd.edu} \quad \href{mailto:ayazdan@google.com}{ayazdan@google.com}}} \\
\textcolor{blue}{\textsf{
	\href{mailto:ksamadi@qti.qualcomm.com}{ksamadi@qti.qualcomm.com} \quad
	\href{mailto:nam.sung.kim@gmail.com}{nam.sung.kim@gmail.com} \quad
	\href{mailto:dburger@microsoft.com}{dburger@microsoft.com} \quad
	\href{mailto:hadi@eng.ucsd.edu}{hadi@eng.ucsd.edu}
}}
}

\usepackage{algorithm2e}
\usepackage{hadianeh}

\usepackage{macros}
\usepackage{nick_names}
\usepackage{results}

\newcommand{\subparagraph}{}
\usepackage{titlesec}
\titleformat{\section}
   {\normalfont\large\bfseries\uppercase}{\thesection.}{3pt}{}
\titleformat{\subsection}
   {\normalfont\large\bfseries}{\thesubsection}{3pt}{}

\usepackage[subtle]{savetrees}
 
\usepackage{natbib}
\begin{document} \sloppy
\maketitle
\thispagestyle{firstpage}
\pagestyle{plain}

\begin{abstract}
Low-power potential of mixed-signal design makes it an alluring option to accelerate Deep Neural Networks (DNNs).
However, mixed-signal circuitry suffers from limited range for information encoding, susceptibility to noise, and Analog to Digital (A/D) conversion overheads.
This paper aims to address these challenges by offering and leveraging the insight that a vector dot-product (the basic operation in DNNs) can be bit-partitioned into groups of spatially parallel low-bitwidth operations, and interleaved across multiple elements of the vectors.
As such, the building blocks of our accelerator become a group of wide, yet low-bitwidth multiply-accumulate units that operate in the analog domain and share a single A/D converter.
The low-bitwidth operation tackles the encoding range limitation and facilitates noise mitigation.
Moreover, we utilize the switched-capacitor design for our bit-level reformulation of DNN operations. 
The proposed switched-capacitor circuitry performs the group multiplications in the charge domain and accumulates the results of the group in its capacitors over multiple cycles.
The capacitive accumulation combined with wide bit-partitioned operations alleviate the need for A/D conversion per operation.
With such mathematical reformulation and its switched-capacitor implementation, we define a 3D-stacked microarchitecture, dubbed \atlass\footnoteremember{name}{\textbf{\atlass}: \textbf{B}it-Partitioned and \textbf{I}nterleaved \textbf{Hi}erarchy of \textbf{W}ide Acceleration through \textbf{E}lectrical Charge}---pronounced Bee Hive---that leverages clustering and hierarchical design to best utilize power-efficiency of the mixed-signal domain and 3D stacking.
For ten DNN benchmarks, \atlass delivers \speedupOverTetris speedup over a leading purely-digital 3D-stacked accelerator \tetris, with a mere of less than 0.5\% accuracy loss achieved by careful treatment of noise, computation error, and various forms of variation.
Compared to RTX~2080~TI with tensor cores and Titan Xp GPUs,  all with 8-bit execution, \atlass offers \perfWattOverRTX and \perfWattOverTitan higher Performance-per-Watt, respectively.
\bihiwe also outperforms other leading digital and analog accelerators in power efficiency.
The results suggest that \atlass is an effective initial step in a road that combines mathematics, circuits, and architecture.
\end{abstract}

\vspace{-1ex}
\section{Introduction}
\label{sec:intro}

Deep Neural Networks (DNNs) are revolutionizing a wide range of services and applications such as language translation~\cite{dnnspeech}, transportation~\cite{dnndriving}, intelligent search~\cite{dnnsearch}, e-commerce~\cite{dnnadvertising}, and medical diagnosis~\cite{dnnmedical}.
These benefits are predicated upon delivery on performance and energy efficiency from hardware platforms.
With the diminishing benefits from general-purpose processors~\cite{dark_silicon:isca, dark_silicon:babak, ccores, dnnoptimizing:fpga:2015}, there is an explosion of digital accelerators for DNNs~\cite{npu:cacm,dadiannao:micro:2014,tetris:asplos:2017, tartan:arxiv:2017, tabla:hpca:2016, cambricon-x:micro:2016,cnvlutin:isca:2016,stripes:micro:2016,dnnweaver:micro:2016, brainwave:hotchips:2017, scnn:isca:2017,yodann:arxiv:2017, eie:isca:2016, eyeriss:isca:2016, eyeriss:jssc:2017, neurocube:isca:16, tpu:isca:2017,diannao:asplos:2014, bitfusion:isca18, snapea:isca:2018, ucnn:isca18, unpu:isscc:2018}.
Mixed-signal acceleration~\cite{isaac:isca:2016, promise:isca18, tsividis1987switched, redeye:isca16, Mour:charge, switched-capacitor, Mour:cifar10, lca:vlsi2017, anpu, zhang201518, passive:switch} is also gaining traction.
Albeit low-power, mixed-signal circuitry suffers from limited range of information encoding, is susceptible to noise, imposes Analog to Digital (A/D) and Digital to Analog (D/A) conversion overheads, and lacks fine-grained control mechanism.
Realizing the full potential of mixed-signal technology requires a balanced design that brings mathematics, architecture, and circuits together.

\begin{figure*}[h]
\begin{minipage}{1\linewidth}
	\centering
	\includegraphics[width=0.88\linewidth]{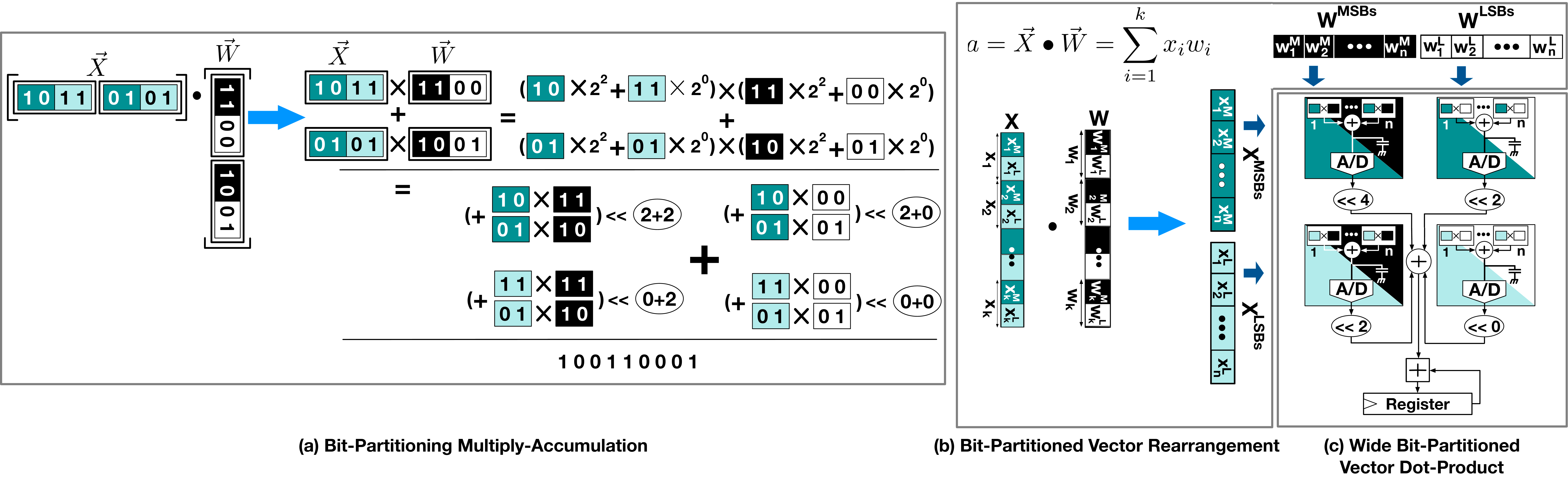} 
	\caption{Wide, interleaved, and bit-partitioned mathematical formulation.}
	\label{fig:compute_model}
\end{minipage}
\vspace{-4ex}
\end{figure*}

This paper sets out to explore this conjunction of areas by inspecting the mathematical foundation of deep neural networks.
Across a wide range of models, the large majority of DNN operations belong to convolution and fully-connected layers~\cite{eyeriss:isca:2016, bitfusion:isca18,isaac:isca:2016}.
Consequently, based on Amdahl's Law, our architecture executes these two types of layers in the mixed-signal domain.
Nevertheless, to maintain generality for the ever-expanding roster of other layers required by modern DNNs, the architecture handles the other layers digitally.
Normally, the convolution and fully-connected layers are broken down into a series of vector dot-products, that generate a scalar and comprise a set of Multiply-Accumulate (MACC) operations.
State-of-the-art digital~\cite{npu:cacm,dadiannao:micro:2014,tetris:asplos:2017, tartan:arxiv:2017, tabla:hpca:2016, cambricon-x:micro:2016,cnvlutin:isca:2016,stripes:micro:2016,dnnweaver:micro:2016, brainwave:hotchips:2017, scnn:isca:2017,yodann:arxiv:2017, eie:isca:2016, eyeriss:isca:2016, eyeriss:jssc:2017, neurocube:isca:16, tpu:isca:2017,diannao:asplos:2014, bitfusion:isca18, snapea:isca:2018, ucnn:isca18, unpu:isscc:2018} and mixed-signal~\cite{isaac:isca:2016, promise:isca18, redeye:isca16, Mour:charge, switched-capacitor, Mour:cifar10, lca:vlsi2017, anpu, tsividis1987switched, gray2001analysis, zhang201518,passive:switch} accelerators use a large array of stand-alone MACC units to perform the necessary computations.
When moving to the mixed-signal domain, this stand-alone arrangement of MACC operations imposes significant overhead in the form of A/D and D/A conversions for each operation.
The root cause is the high cost of converting the operands and outputs of each MACC to and from the analog domain, respectively.

This paper aims to address the aforementioned list of challenges by making the following three contributions.

\niparagraph{(1) This work offers and leverages the insight that the set of MACC operations within a vector dot-product can be partitioned, rearranged, and interleaved at the \emph{bit level} without affecting the mathematical integrity of the vector dot-product.}
Unlike prior work~\cite{ promise:isca18, passive:switch, prime:isca:2016}, this work does not rely on changing the mathematics of the computation to enable mixed-signal acceleration.
Instead, it only rearranges the bit-wise arithmetic calculations to utilize lower bitwidth analog units for higher bitwidth operations.
The key insight is that a binary value can be expressed as the sum of products similar to dot-product, which is also a sum of multiplications ($a = \vec{X}\bullet\vec{W} = \sum_{i} x_i\times w_i$).
Value $b$ can be expressed as $b=\sum_i(2^i\times b_i)$ where $b_i$s are the individual bits or as $b=\sum_{i}(2^{4i}\times bp_i)$, where $bp_i$s are 4-bit partitions for instance.
Our interleaved bit-partitioned arithmetic effectively utilizes the \emph{distributive and associative property} of multiplication and addition at the \emph{bit granularity}.

The proposed model, first, bit-partitions all elements of the two vectors, and then \emph{distributes} the MACC operations of the dot-product over the bit partitions.
Therefore, the lower bitwidth MACC becomes the basic operator that is applied to each bit-partition.  
Then, our mathematical formulation exploits the \emph{associative property} of the multiply and add to group bit-partitions that are at the same significance position.
This significance-based rearrangement enables factoring out the power-of-two multiplicand that signifies the position of the bit-partitions.
The factoring enables performing the wide group-based low-bitwidth MACC operations simultaneously as a spatially parallel operation in the analog domain, while the group shares a \emph{single} A/D convertor.
The power-of-two multiplicand will be applied later in the digital domain  to the accumulated result of the group operation.
To this end, we rearchitect vector dot-product as a series of wide (across multiple elements of the two vectors), interleaved and bit-partitioned arithmetic and re-aggregation.
%
%
Therefore, our reformulation significantly reduces the rate of costly A/D conversion by rearranging the bit-level operations across the elements of  the vector dot-product.
Using low-bitwidth operands for analog MACCs provides a larger headroom between the value encoding levels in the analog domain.
The headroom leads tackles the limited range of encoding and offers higher robustness to noise, an inherent non-ideality in the analog mode. 
Additionally, using lower bitwidth operands reduces the energy/area overhead imposed by A/D and D/A convertors that roughly scales exponentially with the bitwidth of operands.

\niparagraph{(2) At the circuit level, the accelerator is designed using switched-capacitor circuitry that stores the partial results as electric charge over time without conversion to the digital domain at each cycle.}
The low-bitwidth MACCs are performed in charge domain with a set of charge-sharing capacitors.
This design choice lowers the rate of A/D conversion as it implements accumulation as a gradual storage of charge in a set of parallel capacitors.
These capacitors not only aggregate the result of a group of low-bitwidth MACCs, but also enable accumulating results over time.
As such, the architecture enables dividing the longer vectors into shorter sub-vectors that are multiply-accumulated over time with a single group of low-bitwidth MACCs.
The results are accumulated over multiple cycles in the group's capacitors.
Because the capacitors can hold the charge from cycle to cycle, the A/D conversion is not necessary in each cycle.
This reduction in rate of A/D conversion is in addition to the amortized cost of A/D convertors across the bit-partitioned analog MACCs of the group.

\niparagraph{(3) Based on these insights, we devise a clustered 3D-stacked microarchitecture, dubbed \atlass, that provides the capability to integrate copious number of low-bitwidth switched-capacitor MACC units that enables the interleaved bit-partitioned arithmetic.}
The lower energy of mixed-signal computations offers the possibility of integrating a larger number of these units compared to their digital counterpart.
To efficiently utilize the more sizable number of compute units, a higher bandwidth memory subsystem is needed.
Moreover, one of the large sources of energy consumption in DNN acceleration is off-chip DRAM accesses~\cite{ucnn:isca18, bitfusion:isca18, eyeriss:isca:2016}.
%
%
Based on these insights, we devise a clustered architecture for \atlass that leverages 3D-stacking for its higher bandwidth and lower data transfer energy.

Evaluating the carefully balanced design of \atlass with ten DNN benchmarks shows that \atlass delivers \speedupOverTetris speedup over the leading purely digital 3D-stacked DNN accelerator, \tetris~\cite{tetris:asplos:2017}, with only 0.5\% loss in accuracy achieved after mitigating noise, computation error, and Process-Voltage-Temperature (PVT) variations.
With 8-bit execution, \atlass offers \perfWattOverRTX and \perfWattOverTitan higher Performance-per-Watt compared to RTX~2080~TI and Titan Xp, respectively.
With these benefits, this paper marks an initial effort that paves the way for a new shift in DNN acceleration.

\vspace{-1ex}
\section{Wide, Interleaved, and Bit-Partitioned Arithmetic}
\label{sec:compute_model}

%

A key idea of this work is the mathematical insight that enables utilizing low bitwidth mixed-signal units in spatially parallel groups.
This section demonstrates this insight.

\niparagraph{Bit-Level partitioning and interleaving of MACCs.}
To further detail the proposed mathematical reformulation, Figure~\ref{fig:compute_model}(a) delves into the bit-level operations of dot-product on vectors with 2-elements containing 4-bit values.
As illustrated with different colors, each 4-bit element can be written in the form of sum of 2-bit partitions multiplied by powers of 2 (shift).
As discussed, vector dot-product is also a sum of multiplications. 
Therefore, by utilizing the distributive property of addition and multiplication, we can rewrite the vector-dot product in terms of the bit partitions.
However, we also leverage the associativity of the addition and multiplication to group the bit-partitions in the same positions together.
For instance, in Figure~\ref{fig:compute_model}, the black partitions that represent the Most Significant Bits (MSBs) of the $\vec{W}$ vector are multiplied in parallel to the teal\footnote{Color teal in Figure~\ref{fig:compute_model} is the darkest gray in black and white prints.} partitions, representing the MSBs of the $\vec{X}$.
Because of the distributivity of multiplication, the shift amount of (2+2) can be postponed after the bit-partitions are multiply-accumulated.
The different colors of the boxes in Figure~\ref{fig:compute_model} illustrates the interleaved grouping of the bit-partitions.
Each group is a set of spatially parallel bit-partitioned MACC operations that are drawn from different elements of the two vectors.
The low-bitwidth nature of these operations enables execution in the analog domain without the need for A/D conversion for each individual bit-partitioned operation.
As such, our proposed reformulation amortizes the cost of A/D conversion across the bit-partitions of different elements of the vectors as elaborated below.

\niparagraph{Wide, interleaved, and bit-partitioned vector dot-product.}
Figure~\ref{fig:compute_model}(b) illustrates the proposed vector dot-product operation with 4-bit elements that are bit partitioned to 2-bit sub-elements.
For instance, as illustrated, the elements of vector $X$, denoted as $x_i$, are first bit partitioned to $x_i^L$ and $x_i^M$.
The former represents the two Least Significant Bits (LSBs) and the latter represents the Most Significant Bits (MSBs).
Similarly, the elements of vector $W$ are also bit partitioned to the $w_i^L$ and $w_i^M$ sub-elements.
Then, each vector (e.g., $W$) is rearranged into two bit-partitioned sub-vectors, $W^{LSBs}$ and $W^{MSBs}$.
In the current implementations of \atlass architecture, the size of bit-partition is fixed across the entire architecture.
Therefore, the rearrangement is just rewiring the bits to the compute units that imposes modestly minimal overhead (less than 1\%).
Figure~\ref{fig:compute_model} is merely an illustration and there is no need for extra storage or movement of elements.
As depicted with color coding, after the rewiring, $W^{LSBs}$ represents all the least significant bit-partitions from different elements of vector $W$, while the MSBs are rewired in $W^{MSBs}$.
The same rewiring is repeated for the vector $X$.
This rearrangement, puts all the bit-partitions from all the elements of the vectors with the same significance in one group, denoted as $W^{LSBs}$, $W^{MSBs}$, $X^{LSBs}$, $X^{MSBs}$.
Therefore, when a pair of the groups (e.g., $X^{MSBs}$ and $W^{MSBs}$ in Figure~\ref{fig:compute_model}(c)) are multiplied to generate the partial products, (1) the shift amount (``$\ll 4$'' in this case) is the same for all the bit-partitions and (2) the shift can be done after partial products from different sub-elements are accumulated together.

As shown in Figure~\ref{fig:compute_model}(c), the low-bitwidth elements are multiplied together and accumulated in the analog domain.
Accumulation in the digital domain would require an adder tree which is costly compared to the analog accumulation that merely requires connectivity between the multiplier outputs. 
It is only after several analog multiply-accumulations that the results are converted back to digital for shift and aggregation with partial products from the other groups.
The size of the vectors usually exceeds the number of parallel low-bitwidth MACCs, in which case the results need to be accumulated over multiple iterations.
As will be discussed in the next section, the accumulations are performed in two steps.
The first step accumulates the results in the analog domain through charge accumulation in capacitors before A/D convertors (see Figure~\ref{fig:compute_model}(c)).
In the second step, these converted accumulations will be added up in the digital domain using a register.
For this pattern of computation, we are effectively utilizing the \emph{distributive and associative property} of multiplication and addition for dot-product but at the \emph{bit granularity}.
This rearrangement and spatially parallel (i.e., wide) bit-partitioned computation is in contrast with temporally bit-serial digital~\cite{stripes:micro:2016, tartan:arxiv:2017, unpu:isscc:2018, loom:arxiv:2017} and analog~\cite{isaac:isca:2016} DNN accelerators.

The next section describes the architecture of the mixed-signal accelerator that leverages our mathematical reformulation.
This architecture is essentially a collection of the structure that is depicted in Figure~\ref{fig:compute_model}(c). 
The structure is the Mixed-Signal Wide Aggregator (\mswagg) that spatially aggregates the results from its four units as illustrated.
Each of these four units, which are also wide, is a Mixed-Signal Bit-Partitioned MACC (\msbpmacc).
Note that the number of \msbpmaccs in a \mswagg is a function of the bitwidth of the vector elements and the value of bit-partitioning.

\vspace{-1ex}
\section{Mixed-Signal Architecture Design for Wide Bit-Partitioning}
\label{sec:arch}

\begin{figure}
	\centering
	\includegraphics[width=0.8\linewidth]{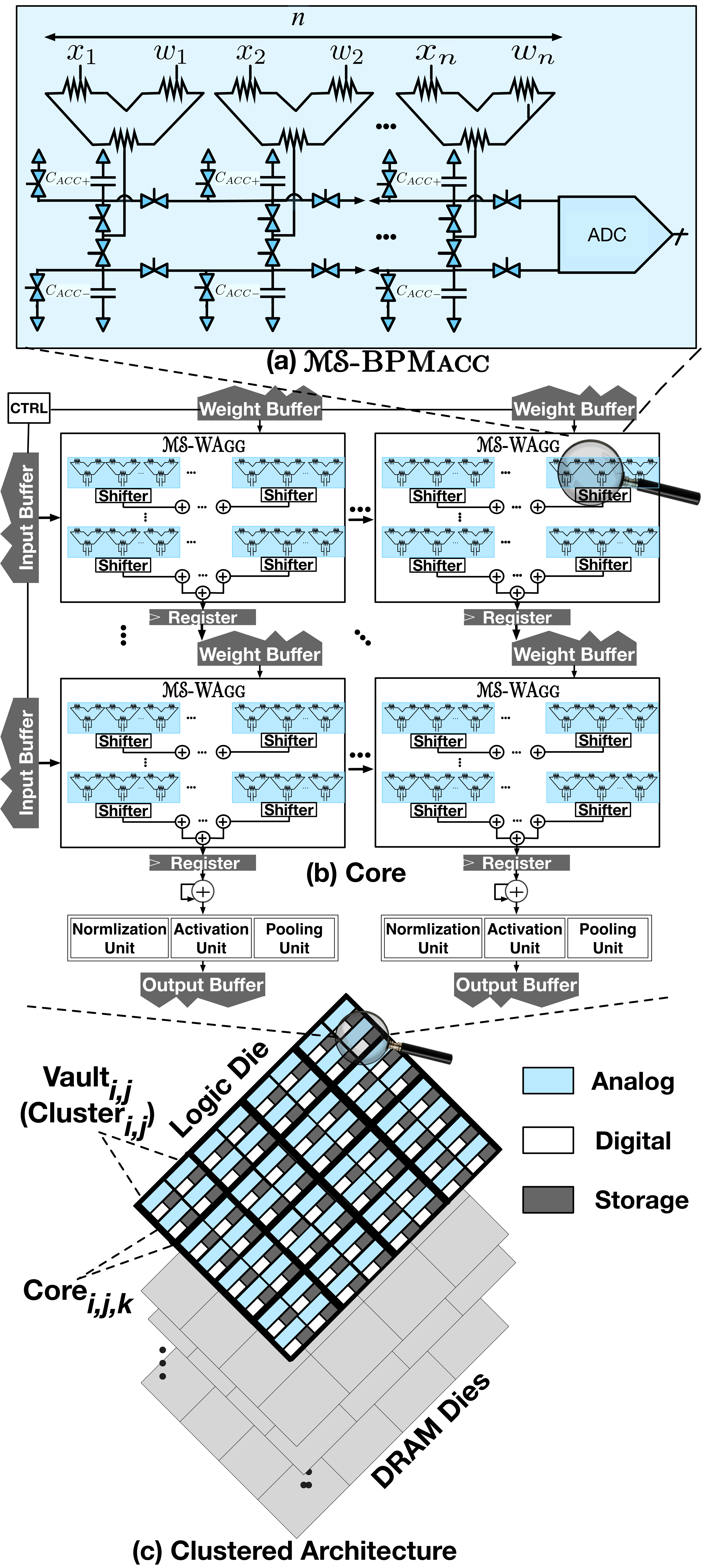}
	\vspace{1pt}
	\caption{Hierarchically clustered architecture of \atlass.}
	\label{fig:arch}
	\vspace{-3.75ex}
\end{figure}

To exploit the aforementioned arithmetic, \atlass comes with a mixed-signal building block that performs wide bit-partitioned vector dot-product.
\atlass then organizes these building blocks in a clustered hierarchical design to efficiently make use of its copious number of parallel low-bitwidth mixed-signal MACC units.
The clustered design is crucial as mixed-signal paradigm enables integrating a larger number of parallel operators than the digital counterpart.
%
%

\subsection{Wide Bit-Partitioned Mixed-Signal MACC}
As Figure~\ref{fig:arch}(a) shows, the building block of \atlass is a collection of low-bitwidth analog MACCs that operate in parallel on sub-elements from the two vectors under dot-product.
This wide structure is dubbed \msbpmacc.
%
%
We design the low-bitwidth MACCs using switched-capacitor circuitry for the following reason.
This design choice lowers the rate of A/D conversion as it implements accumulation as a gradual storage of charge in a set of parallel capacitors.
These capacitors not only aggregate the results of low-bitwidth MACCs, but also enable accumulating results over time.
As such, longer vectors are divided into shorter sub-vectors that are multiply-accumulated over time without the need to convert the intermediate results back to the digital domain.
It is only after processing multiple sub-vectors that the accumulated result is converted to digital, significantly reducing the rate of costly A/D conversions.
As shown in Figure~\ref{fig:arch}(a), each low-bitwidth MACC unit is equipped with its own pair of local capacitors, which perform the accumulation over time across multiple sub-vectors.
As will be discussed in Section~\ref{sec:ms_execute}, the pair is used to handle positive and negative values by accumulating them separately on one or the other capacitor.
After a pre-determined number of private accumulations in the analog domain, the partial results need to be accumulated across the low-bitwidth MACCs.
In that cycle, the transmission gates between the capacitors (Figure~\ref{fig:arch}(a)) connect them and a simple charge sharing between the capacitors yields the accumulated result for the \msbpmacc.
That is when a single A/D conversion is performed, the cost of which is not only amortized across the parallel MACC units but also over time across multiple sub-vectors.

\vspace{-1ex}
\subsection{Mixed-Signal Wide Aggregator}
\msbpmaccs only process low-bitwidth operands; however, they cannot combine these operations to enable higher bit-width dot-products.
A collection of \msbpmaccs can provide this capability as discussed with Figure~\ref{fig:compute_model} in Section~\ref{sec:compute_model}.
This structure is named \mswagg as it is a Mixed Signal Wide Aggregator.
Figure~\ref{fig:arch}(b) depicts a 2D array of a possible \mswagg design, comprising 16 \msbpmaccs that are necessary to perform 8-bit by 8-bit vector dot-product with 2-bit partitioning.
In this case, the number 16 comes from the fact that each of the two 8-bit operands can be partitioned to four 2-bit values.  
Each of the four 2-bit partitions of the multiplicand need to be multiply-accumulated with all the multiplier's four 2-bit partitions.
As discussed in Section~\ref{sec:compute_model}, each \mswagg also performs the necessary shift operations to combine the low-bitwidth results from its 16 \msbpmaccs.
By aggregating the partial results of each \msbpmacc, the \mswagg unit generates a scalar output which is stored on its output register.
As illustrated in Figure~\ref{fig:arch}, a collection of these \mswaggs constitute an accelerator core from which the clustered architecture of \atlass is designed.

\vspace{-1ex}
\subsection{Hierarchically Clustered Architecture}
As discussed in Section~\ref{sec:ms_execute}, the proposed \mswagg consumes $5.4\times$ less energy for a single 8-bit MACC in comparison with a digital logic (1~pJ taken from the \tetris simulator~\cite{tetris:simulator}, which is commensurate with other reports~\cite{li2017caterpillar, 8-bit-mult}).
As such, it is possible to integrate a larger number of mixed-signal compute units on a chip with a given power budget compared to a digital architecture.
To efficiently utilize the larger number of available compute units, a high bandwidth memory substrate is required.
Moreover, one of the large sources of energy consumption in DNN acceleration is off-chip DRAM accesses~\cite{ucnn:isca18, bitfusion:isca18, eyeriss:isca:2016}.
To maximize the benefits of the mixed-signal computation, 3D-stacked memory is an attractive option since it reduces the cost of data accesses and provides a higher bandwidth for data transfer between the on-chip compute and off-chip memory~\cite{tetris:asplos:2017, neurocube:isca:16}.
Based on these insights, we devise a clustered architecture for \atlass with a 3D-stacked memory substrate as shown in Figure~\ref{fig:arch}(c).
The mixed-signal logic die of \atlass is stacked over the DRAM dies with multiple vaults, each of which is connected to the logic die with several through-silicon-via~(TSV)s.
The 3D memory substrate of \atlass is modeled using Micron's Hybrid Memory Cube~(HMC)~\cite{HMC:spec, HMC:vlsi} which has been shown to be a promising technology for DNN acceleration~\cite{tetris:asplos:2017}.
As the results in Section~\ref{sec:results} Figure~\ref{fig:core-design-space} shows, a flat systolic design would result in significant underutilization of the compute resources and bandwidth from 3D stacking.

Therefore, \atlass is a hierarchically clustered architecture that allocates multiple accelerator cores as a cluster to each vault.
Figure~\ref{fig:arch}(b) depicts a single core.
As shown in Figure~\ref{fig:arch}(b), each core is self-sufficient and packs a mixed-signal systolic array of \mswaggs as well as the digital units that perform pooling, activation, normalization, etc.
The mixed-signal array is responsible for the convolutional and fully connected layers.
Generally, wide and interleaved bit-partitioned execution within \mswaggs is orthogonal to the organization of the accelerator architecture.
This paper explores how to embed them and the proposed compute model, within a systolic design and enables end-to-end programmable mixed-signal acceleration for a variety of DNNs.

\niparagraph{Accelerator core.} 
As Figure~\ref{fig:arch}(b) depicts, the first level of hierarchy is the accelerator core and its 2D systolic array that utilizes the \mswaggs.
As depicted, the \code{Input Buffers} and \code{Output Buffers} are shared across the columns and rows, respectively.
Each \mswagg has its own \code{Weight Buffer}. 
This organization is commensurate with other designs and reduces the cost of on-chip data accesses as inputs are reused with multiple filters~\cite{tpu:isca:2017}.
However, what makes our design different is the fact that each buffer needs to supply a sub-vector not a scalar in each cycle to the \mswaggs.
However, the \mswagg generates only a scalar since dot-product generates a scalar output.
The rewiring of the inputs and weights is already done inside the \mswaggs since the size of bit-partitions is fixed.
As such, there is no need to reformat any of inputs, activations, or weights.
As the outputs of \mswaggs flow down the columns, they get accumulated to generate the output activations that are fed to each columns dedicated \code{Normalization}/\code{Activation}/\code{Pooling} \code{Unit}s.
To preserve the accuracy of the DNN model, the intermediate results are stored as 32-bit digital values and intra-column aggregations are performed in the digital mode.

\niparagraph{On-chip data delivery for accelerator cores.}
%
%
To minimize data movement energy and maximally exploit the large degrees of data-reuse offered by DNNs, \atlass uses a statically-scheduled bus that is capable of multicasting/broadcasting data across accelerator cores.
Compared to complex interconnections, the choice of statically-scheduled bus significantly simplifies the hardware by alleviating the need for complicated arbitration logic and FIFOs/buffers required for dynamic routing.
%
%
Moreover, the static schedule enables the \atlass compiler stack to cut the DNN layers across cores while maximizing inter- and intra-core data-reuse.
The static schedule is encoded in the form of data communication instructions (Section~\ref{sec:isa}) that are responsible for (1) fetching data tiles from the 3D-stacked memory and distributing them across cores or (2) writing output tiles back from the cores to the memory.
%

\niparagraph{Parallelizing computations across  accelerator cores.}
Data-movement energy is a significant portion of the overall energy consumption both for digital designs~\cite{tetris:asplos:2017, eyeriss:isca:2016, eyeriss:jssc:2017, bitfusion:isca18, ucnn:isca18, ganax:isca:2018} and analog designs~\cite{promise:isca18, redeye:isca16}.
As such, the \bihiwe clustered architecture (1) divides the computations into tiles that fit within the limited on-chip capacity of the scratchpads that are private for each accelerator core, and (2) \emph{cuts} the tiles of computations across cores to minimize DRAM accesses by maximally utilizing the multicast/broadcasting capabilities of \bihiwe on-chip data delivery network.
To simplify the design of the accelerator cores, the scratchpad buffers are private to each core and the shared data is replicated across multiple cores.
Thus, a single \emph{tile} of data can be read once from the 3D-stacked memory and then be broadcasted/multicasted across cores to reduce DRAM accesses.
%
The cores use double-buffering to hide the latency for memory accesses for subsequent tiles.
%
%
The accelerator cores use \emph{output-stationary} dataflow that minimizes the number of ADC conversions by accumulating results in the charge-domain.
Section~\ref{sec:compiler} discusses the \bihiwe compiler stack that optimizes the \emph{cuts} and \emph{tile} sizes for individual DNN layers.

\vspace{-1ex}
\section{Switched-Capacitor Circuit Design for Bit-Partitioning}
\label{sec:ms_execute}

%

%
\atlass exploits switched-capacitor circuitry~\cite{Mour:charge,  tsividis1987switched, gray2001analysis, passive:switch, zhang201518} for \msbpmacc by implementing MACC operations in the charge-domain rather than using resistive-ladders to compute in current domain~\cite{isaac:isca:2016, anpu, prime:isca:2016}. 
%
Compared to the current-domain approach, switched-capacitors (1) enable result accumulation in the analog domain by storing them as electric charge, eliminating the need for A/D conversion at every cycle, and (2) make multiplications depend only on the ratio of the capacitor sizes rather than their absolute capacitances.
The second property enables reduction of capacitor sizes, improving the energy and area of MACC units as well as making them more resilient to process variation.
%
%
The following discusses the details of the \msbpmacc circuitry.

\vspace{-1ex}
\subsection{Low-Bitwidth Switched-Capacitor MACC}

\begin{figure}
	\centering
	\includegraphics[width=.9\linewidth]{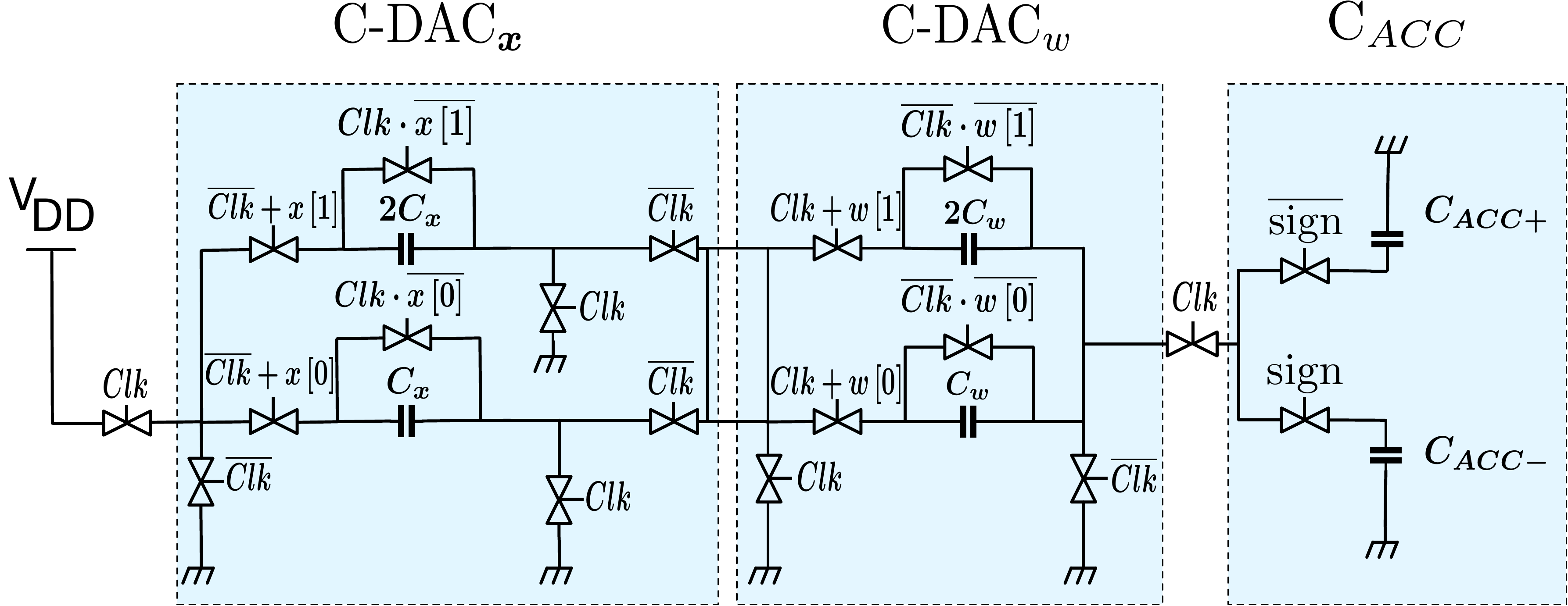}
	\vspace{0.5ex}
	\caption{Low-bitwidth switched-capacitor MACC.}
	\vspace{-6 ex}
	\label{fig:macc}
	\vspace{4ex}
\end{figure}

Figure~\ref{fig:macc} depicts the design of a single 3-bit sign-magnitude MACC. The $x_{s}x_{1}x_{0}$ and $w_{s}w_{1}w_{0}$ denote the bit-partitions operands. 
The result of each MACC operation is retained as electric charge in the accumulating capacitor (\cacc).
%
In addition to \cacc, the MACC unit contains two capacitive Digital-to-Analog Converters, one for inputs (\cdacx ) and one for weights (\cdacw).
The \cdacx and \cdacw convert the 2-bit magnitude of the input and weight to the analog domain as an electric charge proportional to $|x|$ and $|w|$ respectively.
\cdacx and \cdacw are each composed of two capacitors ((\cx, \twocx) and (\cw, \twocw)) which operate in parallel and are combined to convert the operands to analog domain.
Each of these capacitors are controlled by a pair of transmission gates which determine if a capacitor is active or inactive. 
Another set of transmission gates connects the two \cdacs and shares charge when partitions of $x$ and $w$ are multiplied.
The resulting shared charge is stored on either \caccplus or \caccminus depending on the ``sign'' control signal produced by $x_{s}\oplus w_{s}$.
 %
During multiplication, the transmission gates are coordinated by a pair of complimentary non-overlapping clock signals, \clk and \clkb.

\begin{figure}
\begin{minipage}{1\linewidth}
	\centering
	\includegraphics[width=0.75\linewidth]{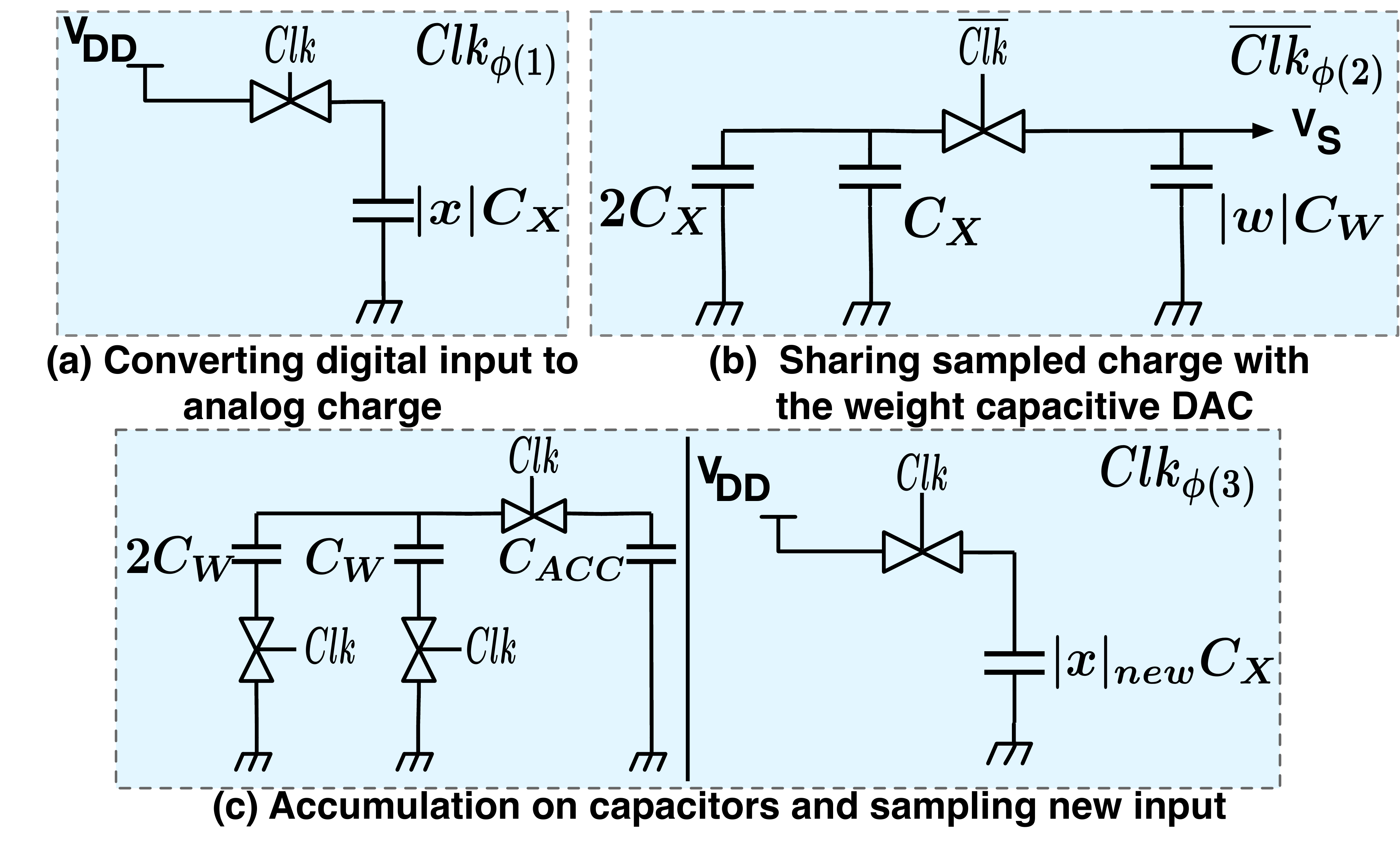}
	\vspace{0.5ex}
	\caption{Charge-domain MACC; phase by phase.}
	\label{fig:timing_diagram}
	\vspace{-4 ex}
\end{minipage}
\end{figure}

\niparagraph{Charge-domain MACC.}
Figure ~\ref{fig:timing_diagram} shows the phase-by-phase process of a MACC and its corresponding active circuits, the phases of which are described below.
%

\niparagraph{$Clk_{\phi(1)}$:}
The first phase (Figure \ref{fig:timing_diagram}(a)) consists of the input capacitive DAC converting digital input ($x$) to a charge proportional to the magnitude of the input $|x|\cx$.
%
%
As a result, the sampled charge ($Q_{sx}$) in \cdacx in the first phase is equal to:
%
\vspace{-1.5ex}
\begin{equation}
    \resizebox{0.3\columnwidth}{!}{
    $Q_{sx} = v_{DD}\times(|X|C_x)$}
\end{equation}
\vspace{-3.5ex}

\niparagraph{$\overline{Clk}_{\phi(2)}$:}
In the second phase (Figure~\ref{fig:timing_diagram}(b)), the multiplication happens via a charge-sharing process between \cdacx and \cdacw.
\cdacw converts the $|w|$ to the charge domain.
At the same time, the \cdacx redistributes its sampled charge ($Q_{sx}$) over all of its capacitors~($3 \times \cx$) as well as the equivalent capacitor of \cdacw.
The voltage ($V_{s}$) at the junction of \cdacx and \cdacw is as follows:
%
\vspace{-1ex}
\begin{equation}
	\resizebox{0.38\columnwidth}{!}{
	$V_{s} = \frac{Q_{sx}}{C_{eq}}= \frac{v_{DD}\times(|X|\cx)}{3\cx + |w|\cw}$}
	\label{eq:redistribute}
	\vspace{-1ex}
\end{equation}
%
Because the sampled charge is shared with the weight capacitors, the stored charge ($Q_{sw}$) on \cdacw is equal to:
%
\vspace{-1.5ex}
\begin{equation}
	\resizebox{0.6\columnwidth}{!}{
	$Q_{sw} = V_s \times |w|\cw = |x|\times|w|\left(\frac{\cw \cx v_{DD}}{3\cx + |w|\cw}\right)$}
\label{eq:weightDAC}
\vspace{-1ex}
\end{equation}
%
Equation~\ref{eq:weightDAC} shows that the stored charge on \cdacw is proportional to $|x|\times|w|$, but includes a non-linearity due to the $|w|$ term in the denominator.
To suppress this non-linearity, $\cx$ and $\cw$ must be chosen such that $3\cx >> |w|\cw$.
 %
 Although this design choice does not completely suppress this non-linearity, it can be mitigated as discussed in Section~\ref{sec:nonideal}.
With this choice, $Q_{sw}$ becomes $Q_{sw} = |x|\times|w|\frac{\cw v_{DD}}{3}$.
%

\niparagraph{$Clk_{\phi(3)}$:} 
In the last phase, (Figure~\ref{fig:timing_diagram}(c)), the charge from multiplication is shared with \cacc for accumulation.
The sign bits ($x_{s}$ and $w_{s}$) determine which of \caccplus or \caccminus is selected for accumulation.
The sampled charge by $|w|\cw$ is then redistributed over the selected \cacc as well as all the capacitors of \cdacw ($=3\cw$).
Theoretically, \cacc must be infinitely larger than $3\cw$ to completely absorb the charge from multiplication.
However, in reality, some charge remains unabsorbed, leading to a pattern of computational error, which is mitigated as discussed in Section~\ref{sec:nonideal}
Ideally, the $V_{ACC}$ voltage on \cacc is:
%
\vspace{-1.5ex}
 \begin{equation}
 \resizebox{0.4\columnwidth}{!}{
 	$V_{ACC} = |x||w|\left(\frac{\cw v_{DD}}{3\times C_{ACC}}\right)$}
 \label{eq:ideal_ACC}
  \vspace{-1ex}
 \end{equation}
 %
While the charge sharing and accumulation happens on \cacc, a new input is fed into \cdacx, starting a new MACC process in a pipelined fashion.
This process repeats for all low-bitwidth MACC units over multiple cycles before one A/D conversion.
\vspace{-1ex}
\subsection{Wide Mixed-Signal Bit-Partitioned MACC} 
\begin{figure}
	\centering
	\includegraphics[width=0.9\linewidth]{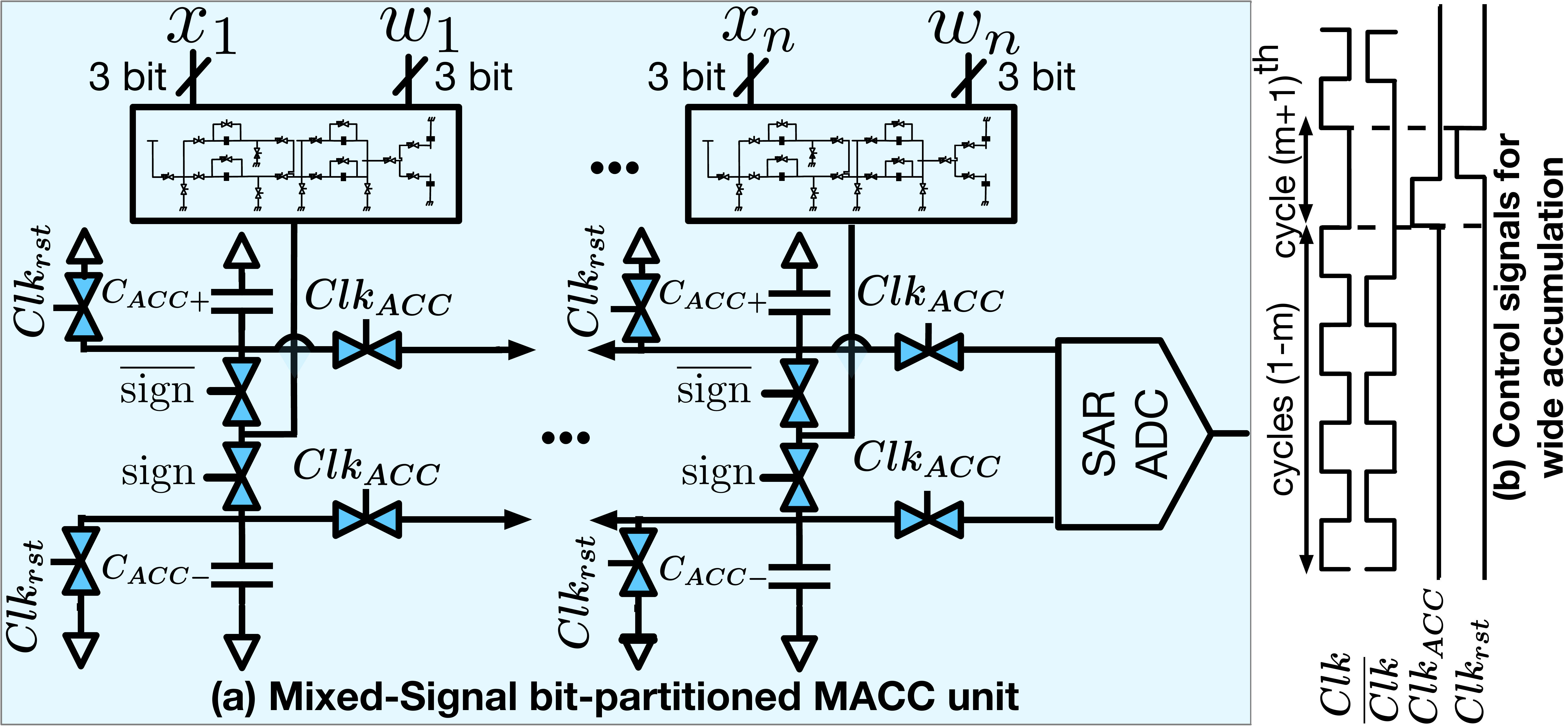}
	\vspace{2pt}
	\caption{Mixed-Signal bit-partitioned MACC unit.}
	\label{fig:msbpmacc}
	\vspace{-2ex}
\end{figure}
Figure~\ref{fig:msbpmacc}(a) depicts an array of $n$ switched-capacitor MACCs, constituting the \msbpmacc unit, which perform operations for $m$ cycles in the analog domain and store the results locally on their \caccs.
Figure~\ref{fig:msbpmacc}(b) depicts the control signals and cycles of operations.
For the \atlass microarchitecture, $m$ and $n$ are selected to $32$ and $8$ based on design space exploration (see Figure~\ref{fig:msbpmacc_design_space}).
Over $m$ cycles, the results of $m \times n$ low-bitwidth MACC operations get accumulated in \caccs, private to each MACC unit.
In cycle $m+1$, the private results get aggregated across all the MACC units within the \msbpmacc.
The \emph{single} A/D converter in the \msbpmacc is responsible for converting the aggregated result, which also starts at cycle $m+1$.

In the first phase of cycle $m+1$, all the $n$ accumulating capacitors which store the positive values (\caccplus) are connected together through a set of transmission gates to share their charge.
Simultaneously, the same process happens for the \caccminus.
\clkacc in Figure~\ref{fig:msbpmacc} is the control signal which connects the \caccs.
The accumulating capacitors (\caccs), are also connected to a Successive Approximation Register (SAR) ADC and share their stored charge with the Sample and Hold block (S\&H) of the ADC.
This (S\&H) block has differential inputs which samples the positive and negative results separately, subtracts them and holds them for the process of A/D conversion.  
In the second phase of the cycle $m+1$, \clkreset connects all the \caccs to ground to clear them for the next iteration of wide, bit-interleaved calculations. 

There is a tradeoff between resolution and sampling rate of ADC,  which also defines its topology.
%
SAR ADC is a better choice when it comes to medium resolution (8-12 bits) and sampling rate (1-500 Mega-Samples/sec).
We choose a 10-bit, 15 Mega-Samples/sec SAR ADC as it strikes the better balance between speed and resolution for \msbpmaccs.
The design space exploration in Figure~\ref{fig:msbpmacc_design_space} shows that this choice makes the grouping of 8 low-bitwidth MACCs optimal for $m=32$ cycles of operation.
The process of A/D conversion takes $m+1$ cycles, pipelined with the sub-vector dot-product.
Table~\ref{tab:power_msbpmacc} shows the energy breakdown within a \msbpmacc that uses 2-bit partitioning.
As shown, performing an 8-bit MACC using the interleaved bit-partitioned arithmetic requires $5.4\times$ less energy than a digital MACC which consumes around 1~pJ~\cite{tetris:asplos:2017}.
\begin{table}[]
\centering
\caption{Energy breakdown for \msbpmacc}
\label{tab:power_msbpmacc}
\scalebox{0.6}{%
\sf
\begin{tabular}{cr}
\hline
\textbf{Units} & \multicolumn{1}{l}{\textbf{Energy (femto Joule)}} \\ \hline
\textbf{1 MACC} & 5.1 fJ \\
\textbf{256 MACCs} & 1,305.6 fJ \\
\textbf{SAR ADC (for 256 MACCs)} & 1,660.0 fJ \\
\textbf{Total Energy} & 1,956.6 fJ \\
\textbf{Total Energy per 2b-2b MACC} & 11.6 fJ \\
\textbf{Total Energy per 8b-8b MACC} & 185.3 fJ \\ \hline
\end{tabular}
}
\vspace{-4ex}
\end{table}

\vspace{-1ex}
\section{Mixed-Signal Non-Idealities and Their Mitigation}
\label{sec:nonideal}
Although analog circuitry offers significant reduction in energy, they might lead to accuracy degradation.
Thus, their error needs to be properly modeled and accounted for.
%
%
Specifically, \msbpmaccs, the main analog component, can be susceptible to (1)~thermal noise, (2)~computational error caused by incomplete charge transfer, and (3) PVT variations.
Traditionally, analog circuit designers mitigate sources of error by just configuring hardware parameters to values which are robust to non-idealities.
Such hardware parameter adjustments require rather significant energy/area overheads that scale linearly with number of modules.
The overheads are acceptable in conventional analog designs since modules are few in numbers.
However, due to the repetitive and scaled-up nature of our design, we need to mitigate these non-idealities in a higher and algorithmic level.
We leverage the training algorithm's inherent mechanism to reduce error (loss) and use mathematical models to represent these non-idealities.
We, then, apply these models during the forward pass to \emph{adjust and fine-tune pre-trained neural models with just a few more epochs} across the chips within a technology node.
%
%
The rest of this section details non-idealities and their modeling.
It, then elaborates on how PVT variations are considered in formulations.

\vspace{-1ex}
\subsection{Thermal Noise}
Thermal noise is an inherent perturbation in analog circuits caused by the thermal agitation of electrons, distorting the main signal.
This noise can be modeled according to a normal distribution, where the ideal voltage deviates relative to a value comprised of the working temperature (T), Boltzmann constant (k),  and capacitor size (C) which produce the deviation $\sigma = \sqrt{kT/C}$. 
Within \atlass, switched-capacitor MACC units are mainly effected by the combined thermal noise resulting from  weights and accumulator capacitors~(\cw and \cacc respectively).
The noise from these capacitors gets accumulated during the $m$ cycles of computation for each individual MACC unit and then gets aggregated  across the $n$ MACC units in \msbpmacc.
By applying the thermal noise equation used for similar MACC units~\cite{passive:switch} to a \msbpmacc unit, the standard deviation at the output is described by Equation ~\ref{eqn:sigma}:
%
\vspace{-6pt}
\begin{equation}
\resizebox{0.75\columnwidth}{!}{
	$\sigma_{ACC} = \sqrt{\frac{kT(\alpha|W_{m-1}| + 3\alpha + 3)}{9\alpha (\alpha +1)^2 C_w} \left(\sum_{i=0}^{m-1} \left(\frac{\alpha}{1+\alpha}\right)^{2i}\right)\times n}$}
	\label{eqn:sigma}
\vspace{-3pt}
\end{equation}
%
%
In the above equation, $\alpha$ is equal to $\frac{\cacc}{3\cw}$.
We apply the effect of thermal noise in the forward propagation of DNN by adding an error tensor to the output of convolutional and fully connected layers.
Having computed the standard deviation of noise for a single \msbpmacc ($\sigma_{ACC}$), each element of the error tensor is sampled from a normal distribution as follows:
\vspace{-6pt}
\begin{equation}
\resizebox{0.43\columnwidth}{!}{
	$\mathcal{N}(\mu=0, \sigma^2=\left(\sigma_{ACC} \times r \times 85\right)^2)$}
\vspace{-3pt}
\end{equation}
In the above equation, $\sigma_{ACC}$ is scaled by $r$ which is the amount of \msbpmacc operations required to generate one element in the output feature map as well as the amount of total bit-shifts applied to each result by \mswagg unit, $85$.
\vspace{-1ex}

\subsection{Computational Error}
Another source of error in \bihiwe's charge-domain computations arises when charge is shared between capacitors during the multiplication and accumulation.
Within each MACC unit, the input capacitors (\cdacx) transfer a sampled charge to the weight capacitors (\cdacw) to produce charge proportional to the multiplication result.
But the resulting charge is subject to error dependent on the ratio of weight and input capacitor sizes ($\beta=C_x/C_w$) as shown in Equation \ref{eq:weightDAC}.
This shared charge in the weight capacitors introduces more error when it is redistributed to the accumulating capacitor (\cacc) which cannot absorb all of the charge, leaving a small portion remaining on the weight capacitors in subsequent cycles.
The ideal voltage ($V_{ACC,Ideal}$) produced after $m$ cycles of multiplication can be derived from Equation~\ref{eq:ideal_ACC} as follows:
%
\vspace{-1ex}
\begin{equation}
\resizebox{0.42\columnwidth}{!}{
$V_{ACC, Ideal}[m] = \sum_{i=1}^m\frac{V\textsubscript{DD}}{9\alpha}W_i X_i$}	
\label{eq:ideal_accumulation}
\vspace{-4pt}
\end{equation}
By considering the computational error from incomplete charge sharing, the actual voltage at the accumulating capacitor after $m$ cycles of MACC operations ($V_{ACC,R}[m]$) becomes:
%
\vspace{-1.5ex}
\begin{equation}
\resizebox{0.7\columnwidth}{!}{
$\frac{3\alpha}{3\alpha+|W_m|}V_{ACC,R}[m-1] +	\frac{W_mX_m \beta}{(3\alpha+|W_m|)(3\beta+|W_m|)}V\textsubscript{DD}$}
\label{eq:real_accumulation}
\vspace{-1ex}
\end{equation}
%
%


%
Computational error is accounted for in the fine-tuning pass by including the multiplicative factors shown in Equation~\ref{eq:real_accumulation} in weights.
During the forward pass, the fine-tuning algorithm decomposes weight tensors in convolutional and fully-connected layers  into groups corresponding to \mswagg configuration and updates the individual weight values ($W_i$)  to new values ($W_i^\prime$) with the computational error in Equation~\ref{eqn:weight_finetune}:
%
\vspace{-1.5ex}
\begin{equation}
\resizebox{0.52\columnwidth}{!}{$
	\begin{split}
	W_i^\prime  = \frac{W_i}{3\alpha + |W_i|}& \frac{\beta V_{DD}}{3\beta + |W_i|}\prod_{j=i+1}^{m-1}\frac{3\alpha}{3\alpha + |W_j|}\\
	 & \forall 0 \leq i \leq m-1
	\end{split}$}
\label{eqn:weight_finetune}
\vspace{-6pt}
\end{equation}
\subsection{Process-Voltage-Temperature Variations}
\niparagraph{Process variations.} We use the sizing of the capacitors to provision and mitigate for the process variations to which the switched-capacitor circuits are generally robust.
The robustness and the mitigation are effective because the capacitors are implemented using a number of smaller unit capacitors with common-centroid layout technique~\cite{analog:layout}.
We, specifically, use the metal-fringe capacitors for MACCs with mismatch of just 1\% standard deviation~\cite{fringe:cap} with the max variation of 6\%~($6\sigma$) which is well below the error margins considered for the computational correctness of \msbpmaccs.

\niparagraph{Temperature variations.} We model the temperature variations by adding a perturbation term to $T$ in Equation~\ref{eqn:sigma} as a gaussian distribution $ \mathcal{N}_T(\mu, \sigma^2)$.
We consider the maximum value of the temperature as 358\degree K which is commensurate with existing practices~\cite{thermal}, and the minimum value as 300\degree K~(This is the peak-to-peak range for the gaussian distribution~($6\sigma$)).

\niparagraph{Voltage variations.} We also model the voltage variation by adding a gaussian distribution to $V_{DD}$ term in Equation~\ref{eqn:weight_finetune}.
Our experiments show that, variations in voltage can be mitigated up to 20\%. 
The extensive amount of vector dot-product operations in DNNs, allows for the minimum and maximum values of the distributions being sampled sufficient amount of times, leading to coverage of the corner cases.

\begin{figure}
	\centering
	\includegraphics[width=1\linewidth]{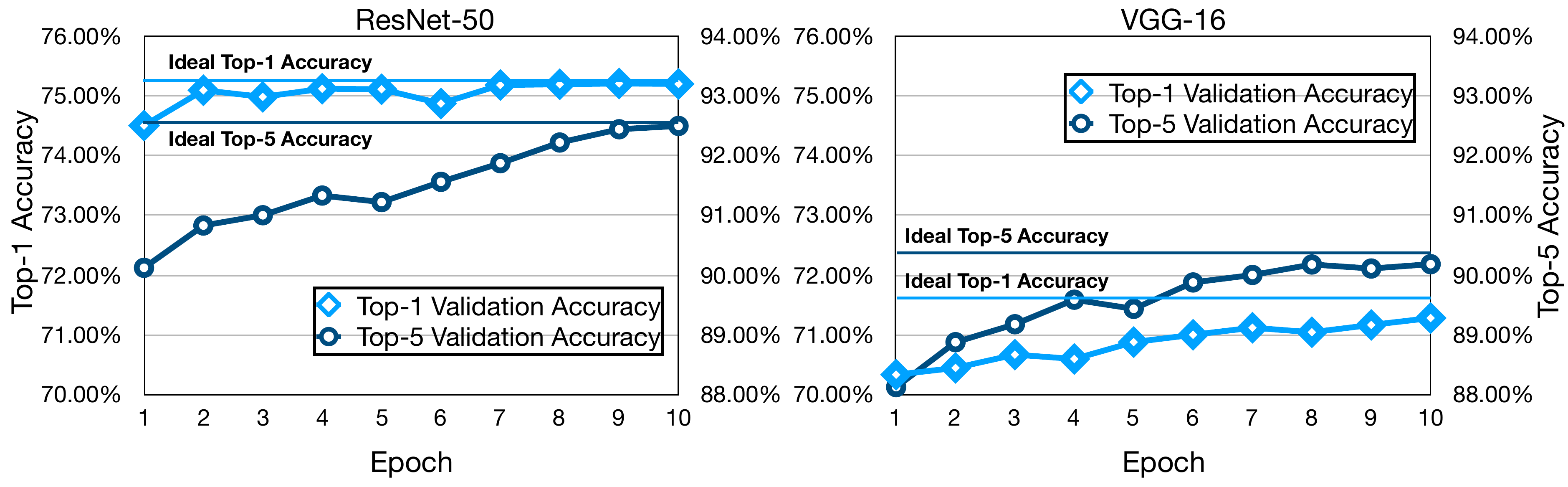}
	\vspace{-1.5ex}
	\caption{\bench{ResNet-50} and \bench{VGG-16} accuracy after fine-tuning.}
	\vspace{-4ex}
	\label{fig:logs_finetuning}
\end{figure}

Atop all these considerations, we use differential signaling for ADCs which attenuates the common-mode fluctuations such as PVT variations.
To show the effectiveness of our techniques, Figure~\ref{fig:logs_finetuning} plots the result of fine-tuning process of two benchmarks, \bench{ResNet-50} and \bench{VGG-16} for ten epochs.
Table~\ref{tab:retraining} reports the summary of accuracy trends for all the benchmarks, which achieve less than 0.5\% loss.
As Figure~\ref{fig:logs_finetuning} shows, the fine-tuning pass compensates the initial loss (0.73\% for top-1 and 2.41\% for top-5) to only 0.04\% for top-1 and 0.02\% for top-5.
\bench{VGG-16} is slightly different and reduces the initial loss (1.16\% for top-1 and 2.24\% for top-5) to less than 0.18\% for top-1 and 0.13\% for top-5 validation accuracy.
The trends are similar for other benchmarks and omitted due to space constraints.
%

%


\vspace{-0.5ex}
\section{{\large\textbf{\atlass}} Compiler Stack}
\label{sec:compiler}

\begin{figure}[b]
	\vspace{-3ex}
	\centering
	\includegraphics[width=0.75\linewidth]{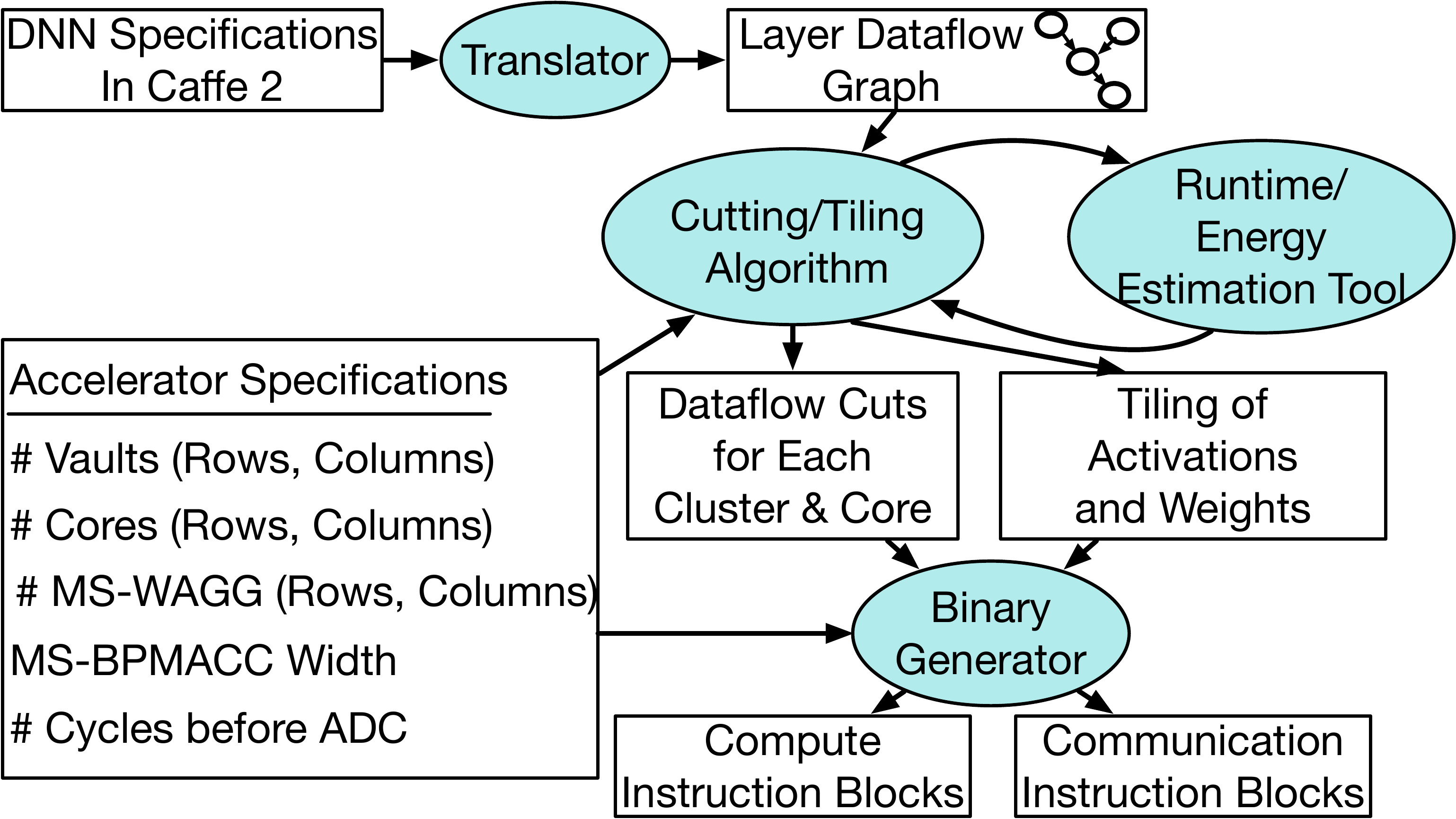}
	\vspace{0.5ex}
	\caption{\atlass compilation stack.}
	\vspace{-2ex}
	\label{fig:compilation-flow}
\end{figure}


%
As Figure~\ref{fig:compilation-flow} shows, DNNs are compiled to \atlass through a multi-stage process beginning with a Caffe2~\cite{caffe2} DNN specification file.
The high-level specification provided in the Caffe2 file is translated to a layer DataFlow Graph (DFG) that preserves the structure of the network.
The DFG goes through an algorithm that cuts the DFG and tiles the data to map the DNN computations to the accelerator clusters and cores.
The tiling also aims to minimize the transfer of model parameters to limited on-chip scratchpads on the logic die from the 3D-stacked DRAM, while maximizing the utilization of the compute resources.
In addition to the DFG, the cutting/titling algorithm takes in the architectural specification of the \atlass.
These specifications include the organizations and configurations (\# rows, \#columns) of the clusters, vaults, and cores as well as details of the \msbpmaccs.
To identify the best cuts and tilings, the cutting/tiling algorithm exhaustively searches the space of possibilities, which is enabled through an estimation tool.
The tool estimates the total energy consumption and runtime for each cuts/tiles pair which represent the data movement and resource utilization in \bihiwe.
Estimation is viable, as the DFG does not change, there is no hardware managed cache, and the accelerator architecture is fixed during execution. 
Thus, there are no irregularities that can hinder estimation.
Algorithm~1 depicts the cutting/tiling procedure.
When cuts and tiles are determined, the compiler generates the binary code that contains the communication and computation instruction blocks.
As commensurate with state-of-the-art accelerators~\cite{tetris:asplos:2017, bitfusion:isca18, eyeriss:isca:2016, neurocube:isca:16, dnnweaver:micro:2016}, all the instructions are statically scheduled.
We extend the static scheduling to cluster coordination, data communication and transfer.
%
\begin{algorithm}
\vspace{-3ex}
\fontsize{7.5}{8.5}\selectfont
\tline
\vspace{2pt}
   	Initialize $cut_{opt}[N] \gets \emptyset$\\
   	Initialize $tiling_{opt}[N] \gets \emptyset$\\
	\For{ $layer_i \in \text{DFG}_{DNN}$ }
	{
		\vspace{0.5ex}
		$s_{opt}\gets \infty$\\
		\For {$tiling_{i,j}\in layer_i$ }
		{
			\For {$cut_{i,j,k} \in tiling_{i,j}$ }
			{
				$(\text{runtime}_{i,j,k},~\text{energy}_{i,j,k}) \gets \text{EstimationTool}(tiling_{i,j}, cut_{i,j,k})$\\
				$s_{i,j,k} \gets \text{runtime}_{i,j,k} \times \text{energy}_{i,j,k}$\\
				\vspace{0.5ex}
				\uIf {$s_{i,j,k} < s_{opt}$}
				{
					$cut_{opt}[i] \gets cut_{i,j,k}$\\
					$tiling_{opt}[i] \gets tiling_{i,j}$\\
				}
			}	
		}
	}
	\Return $cut_{opt}$, $tiling_{opt}$
\vspace{1.5ex}
\bline
\text{\textsf{\textbf{\small Algorithm 1: Cutting/tiling algorithm for clustered acceleration.\label{alg:clustering}
}}}
\vspace{-8ex}

\end{algorithm}

\vspace{0ex}
\section{{\large\textbf{\atlass}} Instruction Set}
\label{sec:isa}

%



The \atlass ISA exposes the following unique properties of its architecture to the software: (1) efficient \textit{mixed-signal execution} using bit-partitioned \mswagg and \textit{capacitive accumulation}, and (2) \textit{clustered architecture}, that takes advantage of the power efficiency of mixed-signal acceleration to scale-up the number of \mswaggs in \atlass.
As such, \atlass uses a block-structured ISA that segregates the execution of the DNN into (1) data \code{communication instruction blocks} that accesses \code{tiles} of data from the 3D-stacked memory and populates the on-chip scratchpads (Input Buffer/Weight Buffer/Output Buffer in Figure~\ref{fig:arch}), and (2) \code{compute instruction blocks} each of which consumes the tile of data produced by a corresponding communication instruction block and produces an output tile.
The \atlass compiler stack statically assigns communication and compute instruction blocks to accelerator clusters, shifting the complexity from hardware to the compiler.
By splitting the data transfer and on-chip data processing into separate instructions, the \atlass ISA enables software pipelining between clusters and allows the memory accesses to run ahead and fetch data for the next tile while processing the current tile.

\niparagraph{Compute instruction block.}
A block of compute instructions expresses the entire computation to produce a single tile in an accelerator core.
Further, the compute block governs how the input data for a DNN layer is bit-partitioned and distributed across wide aggregators within a single core.
As such, the compiler has complete control over the read/write accesses to on-chip scratchpads, A/D and D/A conversion, and execution using the \mswaggs and digital blocks in an accelerator core.
%
%
The granularity of bit-partitioning and charge-based accumulation is determined for each microarchitectural implementation based on the technology node and circuit design paradigm.
As such, to support different technology nodes and design styles and allow extensions to the architecture, the \atlass ISA encodes the bit-partitioning and accumulation cycles.
However, we need to explore the design space to find the optimal design choice for each combination of technology node and circuits (Section~\ref{sec:eval}).

%

\niparagraph{Communication instruction block.}
The key challenge when scaling up the design is to minimize data-movement while parallelizing the execution of the DNN across the on-chip compute resources.
To simplify the hardware, \atlass instruction set captures the static schedule of data movement as a series of \code{communication instruction block}s.
Static scheduling is possible as the topology of the DNN does not change during inference and the order of layers and neurons is known statically.
The \atlass compiler stack assigns the communication blocks to the cores according to the order of the layers.
This static ordering enables \atlass to use a simple statically scheduled bus instead of a more complex interconnection.
%

To maximize energy efficiency, it is imperative to exploit the high degree of data-reuse offered by DNNs.
To exploit data-reuse when parallelizing computations across cores of the \atlass architecture, the communication instructions support broadcasting/multicasting to distribute the same data across multiple cores, minimizing off-chip memory accesses.
Once a communication block writes a tile of data to the on-chip scratchpads, it can be reused over multiple compute blocks to exploit temporal data locality within a single accelerator core.


\vspace{-1ex}
\section{Evaluation}
\label{sec:eval}
\vspace{-1ex}
\subsection{Methodology}
\label{sec:method}
\vspace{-1ex}
\begin{table}[]
\centering
\vspace{-1ex}
\caption{Evaluated benchmarked DNNs}
\label{tab:benchmarks}
\scalebox{0.5}{%
\sf
\begin{tabular}{@{}ccccrr@{}}
\toprule
\textbf{DNN} & \textbf{Type} & \textbf{Domain} & \textbf{Dataset} & \multicolumn{1}{c}{\textbf{Multiply-Adds}} & \multicolumn{1}{c}{\textbf{Model Weights}} \\ \midrule
\textbf{AlexNet~\cite{alexnet}} & CNN & Image Classification & Imagenet~\cite{imagenet} & 2,678 MOps & 56.1 MBytes \\
\textbf{CIFAR-10}~\cite{vgg, qnn:arxiv:2016} & CNN & Image Classification & CIFAR-10~\cite{cifar10} & 617 MOps & 13.4 MBytes \\
\textbf{GoogLeNet}~\cite{googlenet} & CNN & Image Classification & Imagenet & 1,502 MOps & 13.5 MBytes \\
\textbf{ResNet-18}~\cite{resnet} & CNN & Image Classification & Imagenet & 4,269 MOps & 11.1 MBytes \\
\textbf{ResNet-50}~\cite{resnet} & CNN & Image Classification & Imagenet & 8,030 MOps & 24.4 MBytes \\
\textbf{VGG-16}~\cite{vgg} & CNN & Object Recognition & Imagenet & 31 GOps & 131.6 MBytes \\
\textbf{VGG-19}~\cite{vgg} & CNN & Object Recognition & Imagenet & 39 GOps & 137.3 MBytes \\
\textbf{YOLOv3}~\cite{yolov3} & CNN & Object Recognition & Imagenet & 19 GOps & 39.8 MBytes \\
\textbf{PTB-RNN}~\cite{qnn:arxiv:2016} & RNN & Language Modeling & Penn TreeBank~\cite{penn-treebank} & 17 MOps & 16 MBytes \\
\textbf{PTB-LSTM}~\cite{lstm} & RNN & Language Modeling & Penn TreeBank & 13 MOps & 12.3 MBytes \\ \bottomrule
\end{tabular}
}
\vspace{-4ex}
\end{table}

\begin{table}[!hb]
\vspace{-1ex}
\centering
\caption{\atlass and baselines platforms}
\label{tab:platforms}
\scalebox{0.52}{%
\sf
\begin{tabular}{crrcrr}
\hline
\textbf{Parameters} & \multicolumn{2}{c}{\textbf{ASIC}} & \textbf{Parameters} & \multicolumn{2}{c}{\textbf{GPU}} \\ \hline
\textbf{Chip} & \multicolumn{1}{c}{\textbf{\atlass}} & \multicolumn{1}{c|}{\textbf{\tetris}} & \textbf{Chip} & \multicolumn{1}{c}{\textbf{RTX 2080 TI}} & \multicolumn{1}{c}{\textbf{Titan Xp}} \\
\textbf{MACCs} & 16,384 & \multicolumn{1}{r|}{3,136} & \textbf{Tensore Cores} & 544 & --- \\
\textbf{On-chip Memory} & 9216 KB & \multicolumn{1}{r|}{3698 KB} & \textbf{Memory} & 11 GB~(GDDR6) & 12 GB~(GDDR5X) \\
\multirow{2}{*}{\textbf{\begin{tabular}[c]{@{}c@{}}Chip Area\\ ($mm^2$)\end{tabular}}} & \multirow{2}{*}{122.3} & \multicolumn{1}{r|}{\multirow{2}{*}{56}} & \textbf{Chip Area~($mm^2$)} & 754 & 471 \\
 &  & \multicolumn{1}{r|}{} & \textbf{Total Dissipation Power} & 250 W & 250 W \\
\textbf{Frequency} & 500 Mhz & \multicolumn{1}{r|}{500 Mhz} & \textbf{Frequency} & 1545 Mhz & 1531 Mhz \\
\textbf{Technology} & 45 nm & \multicolumn{1}{r|}{45 nm} & \textbf{Technology} & 12 nm & 16 nm \\ \bottomrule
\end{tabular}
}
\vspace{-2.5ex}
\end{table}
%

%
\niparagraph{Benchmarks.}
We use ten diverse CNN and RNN models to evaluate \atlass, described in Table~\ref{tab:benchmarks} that perform image classification, real-time object detection (\bench{YOLOv3}), and character-level (\bench{PTB-RNN}) and word-level (\bench{PTB-LSTM}) language modeling.
This set of benchmarks includes medium to large scale models (from 11.1 MBytes to 137.3 MBytes) and variety of multiply-add operations (from 13 Million to 39 Billion). 
%

%
%
%
%
%
%
%

\niparagraph{Simulation infrastructure.}
We develop a cycle-accurate simulator and a compiler for \atlass that takes in a \code{caffe-2} specification of the DNN, finds the optimum tiling and cutting for each layer, and maps it to \atlass architecture.
%
The simulator executes each of the optimized network using the \atlass architecture model and reports the total runtime and energy.

\niparagraph{\tetris comparison.}
We compare \atlass with \tetris, a state-of-the-art fully-digital 3D-stacked dataflow accelerator.
We match the on-chip power dissipation of \atlass and \tetris and compare the total runtime and energy, including energy for DRAM accesses.
We also perform an iso-area comparison and scale up original \tetris with 16 vaults to 36 vaults to match its area to \atlass's.
The baseline \tetris supports 16-bit execution while \atlass supports 8-bit.
For fairness, we modify the open-source \tetris simulator~\cite{tetris:simulator} and proportionally scale its runtime and energy.
\atlass supports 8-bit operands since this representation has virtually no impact by itself on the final accuracy of the DNNs~\cite{qnn:arxiv:2016, dorefa:arxiv:2016, wrpn, li2016ternary, zhang2018lq}.

\niparagraph{GPU comparison.}
We also compare \bihiwe to two Nvidia GPUs (i.e., RTX 2080 TI and Titan Xp) based on Turing and Pascal architecture respectively, listed in Table~\ref{tab:platforms}.
RTX 2080 TI's Turing architecture provides tensor cores which are specialized hardware for deep learning inference.
We use 8-bit on GPUs using Nvidia's own \code{TensorRT 5.1}~\cite{tensorrt} library compiled with the optimized \code{cuDNN 7.5} and \code{CUDA 10.1}.
For each DNN benchmark, we perform 1,000 warmup iterations and report the average runtime across 10,000 iterations.

\niparagraph{Comparison with other recent accelerators.}
We also compare \bihiwe to Google TPU~\cite{tpu:isca:2017}, mixed-signal CMOS RedEye~\cite{redeye:isca16}, and two analog memristive accelerators.
All the comparisons are in 8-bits.
The original designs~\cite{isaac:isca:2016,pipelayer:hpca:17} use 16-bits.
%
%
Scaling from 16-bit to 8-bit execution for memristive designs would optimistically provide a $4\times$ increase in efficiency.
%
%
%

%
\niparagraph{Energy and area measurement.}
All hardware modelings are performed using \code{FreePDK 45-nm} standard cell library~\cite{freepdk}.
We implement the switched-capacitor MACCs in \code{Cadence Analog Design Environment V6.1.3} and use \code{Spectre SPICE V6.1.3} to model the system.
We then, use \code{Layout XL} of Cadence to lay out the MACC units and extract the energy/area.
The ADC's energy/area are obtained from \cite{survey}.
%
Based on the \msbpmacc configuration, we use the ADC architecture from~\cite{cicc18:ADC}.
%
%
%

We implement all digital blocks of \atlass, including adders, shifters, interconnection, and accumulators in Verilog RTL and used \code{Synopsys Design Compiler (L-2016.03-SP5)} to synthesize them and measure their energy and area.
For on-chip SRAM buffers, we use \code{CACTI-P} ~\cite{cactip} to measure the energy and area of the memory blocks.
The 3D-stacked DRAM architecture is based on HMC stack \cite{HMC:spec, HMC:vlsi}, the same as \tetris, and the bandwidth and access energy are adopted form that work.
%
%
%

\niparagraph{Error modeling.}
For error modeling, we use \code{Spectre SPICE V6.1.3} to extract the noise behavior of MACCs via circuit simulations.
Thermal noise, computational error, and PVT variations are considered based on details in Section~\ref{sec:nonideal}.
We implement the extracted hardware error models and the corresponding mathematical modelings using \code{PyTorch v1.0.1}~\cite{pytorch} and integrate them into \code{Neural Network Distiller v0.3} framework~\cite{distiller} for a fine-tuning pass over the evaluated benchmarks. 
%

\vspace{-1ex}
\subsection{Experimental Results}
\label{sec:results}
%

\subsubsection{Comparison with \tetris}
\begin{figure}
\vspace{0ex}
	\centering
	\includegraphics[width=0.9\linewidth]{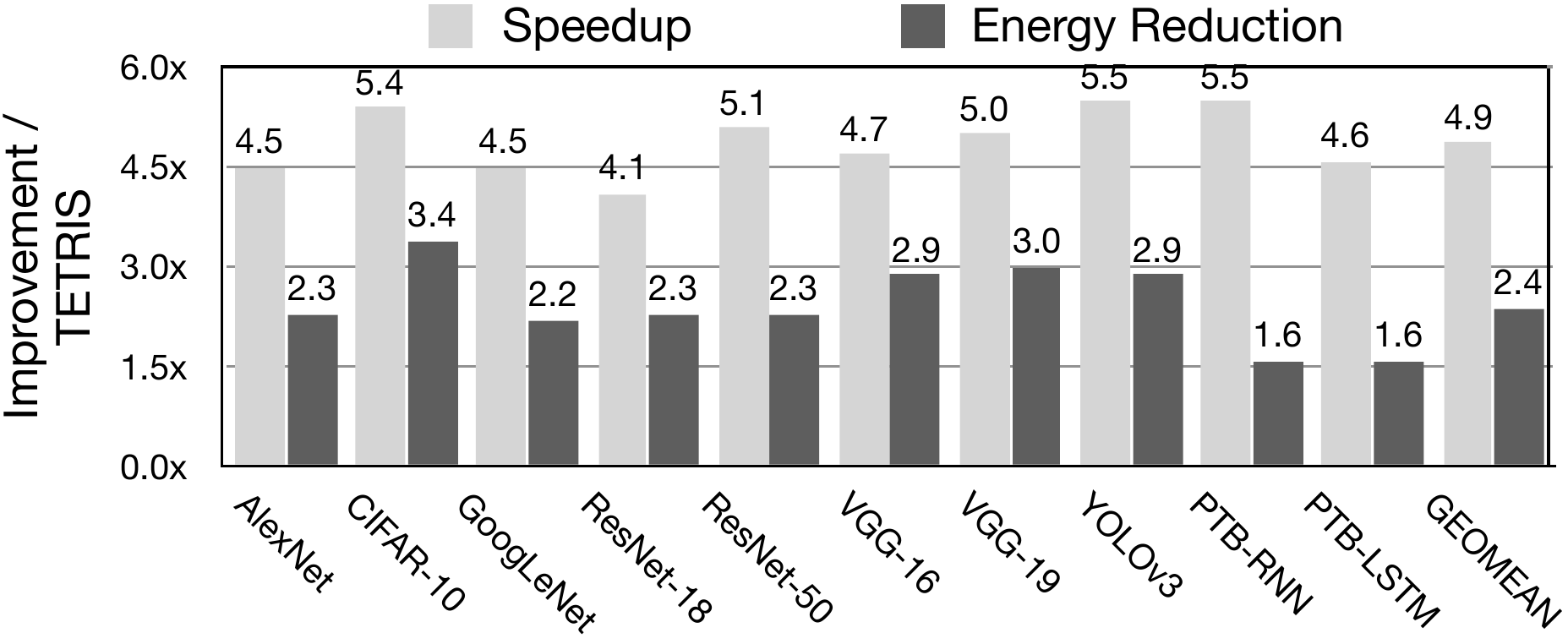}
	\caption{Speedup and energy improvement over \tetris.}
	\label{fig:tetris_comparison}
	\vspace{-4ex}
\end{figure}

\niparagraph{Iso-power performance and energy comparison.} Figure \ref{fig:tetris_comparison} shows the performance and energy reduction of \atlass over \tetris under the same on-chip power budget.
On average, \atlass delivers a \speedupOverTetris speedup over \tetris.
This significant improvement is attributed to the use of wide mixed-signal \msbpmaccs in \atlass as opposed to PEs in \tetris.
The wide bit-partitioned mixed-signal design of \msbpmacc in \atlass enables us to cram $\approx$ 5$\times$ more compute units within the same power budget as \tetris .  
The highest speedup is observed in \bench{YOLOv3} and \bench{PTB-RNN}, where their networks' configurations favor the wide vectorized execution in \bihiwe by better utilizing compute resources.
The lowest speedup is observed in \bench{ResNet-18}, since its relatively small size leads to under-utilization of compute resources in \bihiwe.
%
%
%

Figure \ref{fig:tetris_comparison} demonstrates the total energy reduction for \atlass across the evaluated benchmarks as compared to \tetris.
On average, \atlass yields \energyOverTetris energy reduction over \tetris, including energy for DRAM accesses, while consuming the same on-chip power as \tetris.
\bench{CIFAR-10} enjoys the highest energy reduction, since \bihiwe is able to take advantage of \bench{CIFAR-10}'s smaller memory footprint to maximize on-chip data reuse and reduce DRAM accesses.
The lowest energy reduction is observed in RNN benchmarks, \bench{PTB-RNN} and \bench{PTB-LSTM} since the matrix-vector operations in these benchmarks require a significant number of memory accesses, diminishing the benefits from mixed-signal computations.

%

\niparagraph{Energy breakdown.}
\begin{figure}[!hb]
\vspace{-1.5ex}
	\centering
	\includegraphics[width=1\linewidth]{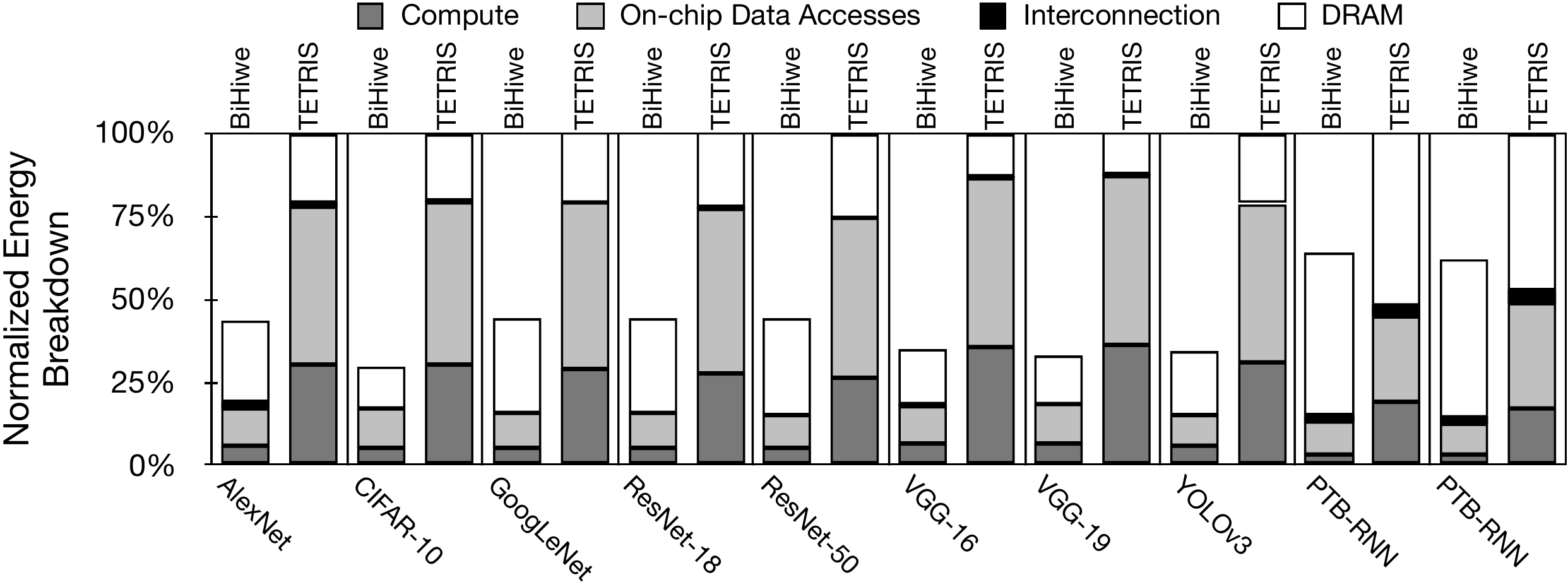}
	\caption{Energy breakdown of \atlass and \tetris.}
	\label{fig:energy_breakdown}
	\vspace{-1ex}
\end{figure}
\begin{figure}[]
\vspace{-1ex}
	\centering
	\includegraphics[width=0.9\linewidth]{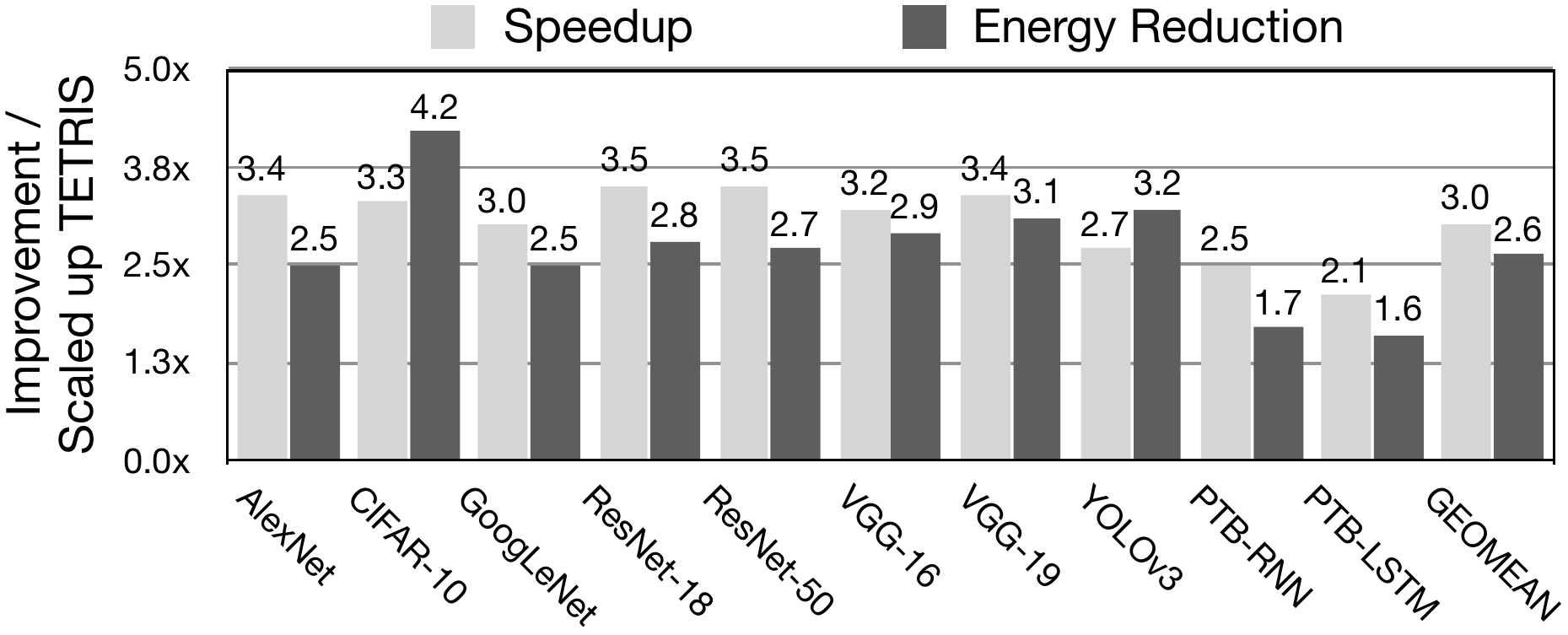}
	\caption{Iso-area comparison with \tetris.}
	\label{fig:tetris_comparison_iso_area}
	\vspace{-3ex}
\end{figure}
Figure~\ref{fig:energy_breakdown} shows the energy breakdown normalized to \tetris.
Energy breakdown is reported across four major architectural components: (1) on-chip compute units, (2) on-chip memory (buffers and register file), (3) interconnect, and (4) 3D-stacked DRAM.
DRAM accesses account for the highest portion of the energy in \atlass, since \atlass significantly reduces the on-chip compute energy. 
While \atlass has a significantly larger number of compute resources compared to \tetris, the number of DRAM accesses remain almost the same.
This is because the statically-scheduled bus allows data to be multicasted/broadcasted across multiple cores in \atlass without significantly increasing the number of DRAM accesses.
Furthermore, the statically-scheduled bus offers the \atlass compiler stack the freedom to optimize partitioning the computations across cores.
Most layers in the benchmarks benefit for partitioning the different inputs in a single batch (batch size is 16) across \atlass cores and broadcasting weights, which is not explored in \tetris.
%
As a result, these networks have lower DRAM accesses.
The breakdown of energy consumption varies with the type of computations required by the DNN as well as the degree of data-reuse.
Benchmarks \bench{PTB-RNN} and \bench{PTB-LSTM} are recurrent neural networks that perform large matrix-vector operations and require significant DRAM accesses for weights.
Therefore, \bench{PTB-RNN} and \bench{PTB-LSTM} use more energy for DRAM accesses compared to other benchmarks.

Unlike the fully-digital PEs in \tetris that perform a single operation in a cycle, \atlass uses \mswaggs which perform wide vectorized operations--crucial in \atlass to amortize the high cost of ADCs.
As shown in Table \ref{tab:power_msbpmacc}, each MACC operation in \atlass consumes 5.4$\times$ less energy compared to \tetris.
The output-stationary dataflow enabled by capacitive accumulation in addition to the systolic organization of \mswaggs in each core of \atlass which eliminates the need for register files unlike \tetris, leads to 4.4$\times$ reduction for on-chip data movement on average.

\niparagraph{Iso-area comparison with \tetris.}
We compare the total runtime and energy of \bihiwe with a scaled up version of \tetris which matches \bihiwe'area.
Figure~\ref{fig:tetris_comparison_iso_area} shows the results for the workloads.
Scaling-up the compute resources in \tetris by 2.25$\times$ to match the chip-area of \bihiwe results in a sub-linear increase in performance by $\approx60\%$.
This improvement in performance comes at a cost of reduced energy-efficiency due to an increase in memory accesses to feed the additional compute resources.
The trends in speedup and energy-reduction remain the same as iso-power comparison, with the exception of \bench{ResNet-18}, which now sees resource underutilization in \tetris after scaling up number of compute resources.
%
 
\subsubsection{Comparison to GPUs}
\begin{figure}
\vspace{0ex}
	\centering
	\includegraphics[width=0.9\linewidth]{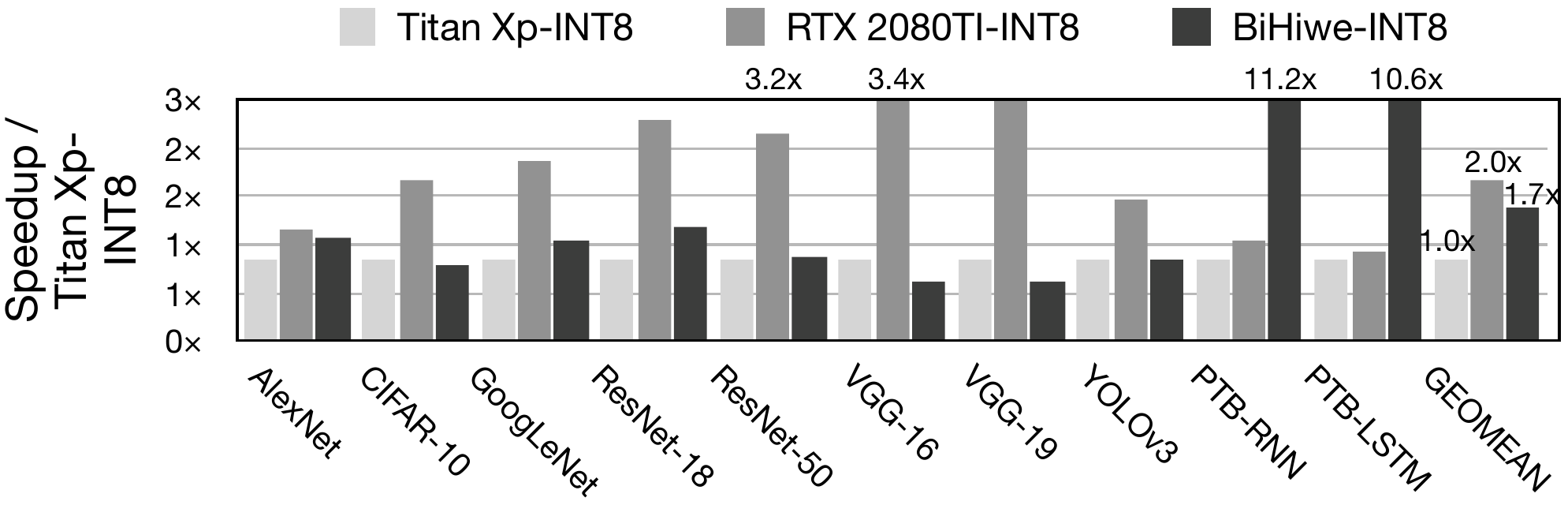}
	\caption{Performance comparison to GPUs.}
	\label{fig:GPU_comparison}
	\vspace{-4.5ex}
\end{figure}
Figure~\ref{fig:GPU_comparison} compares performance of \atlass with Titan~Xp and RTX~2080~TI.
RTX~2080~TI is based on Nvidia's latest architecture, Turing.
For a fair comparison, we enable vectorized 8-bit operations and optimized GPU compilations.
The results are normalized to Titan~Xp. 
%
%
\atlass, on average, yields 70\% speedup over Titan~Xp GPU and performs 15\% slower than RTX~2080~TI.
Convolutional networks require large amount of matrix-matrix multiplications that are well-suited for tensor cores, leading to RTX~2080~TI's outperformance on both \bihiwe and Titan~Xp.
\bench{VGG-16} and \bench{VGG-19} see the maximum benefits.
However, \atlass outperforms RTX~2080~TI GPU in \bench{PTB-RNN} and \bench{PTB-LSTM} with 11.2$\times$ and 10.6$\times$, respectively.
These RNN networks require matrix-vector multiplications, which is particularly suitable for the wide vectorized operations supported in \atlass's \mswaggs--not the best case for tensor cores.
In terms of performance-per-Watt, \atlass outperforms both Titan~Xp and RTX~2080~TI GPUs by a large margin, \perfWattOverTitan and \perfWattOverRTX, respectively.

\subsubsection{Comparison with Other Accelerators}

\begin{figure}[!hb]
\vspace{-1.5ex}
	\centering
	\includegraphics[width=0.7\linewidth]{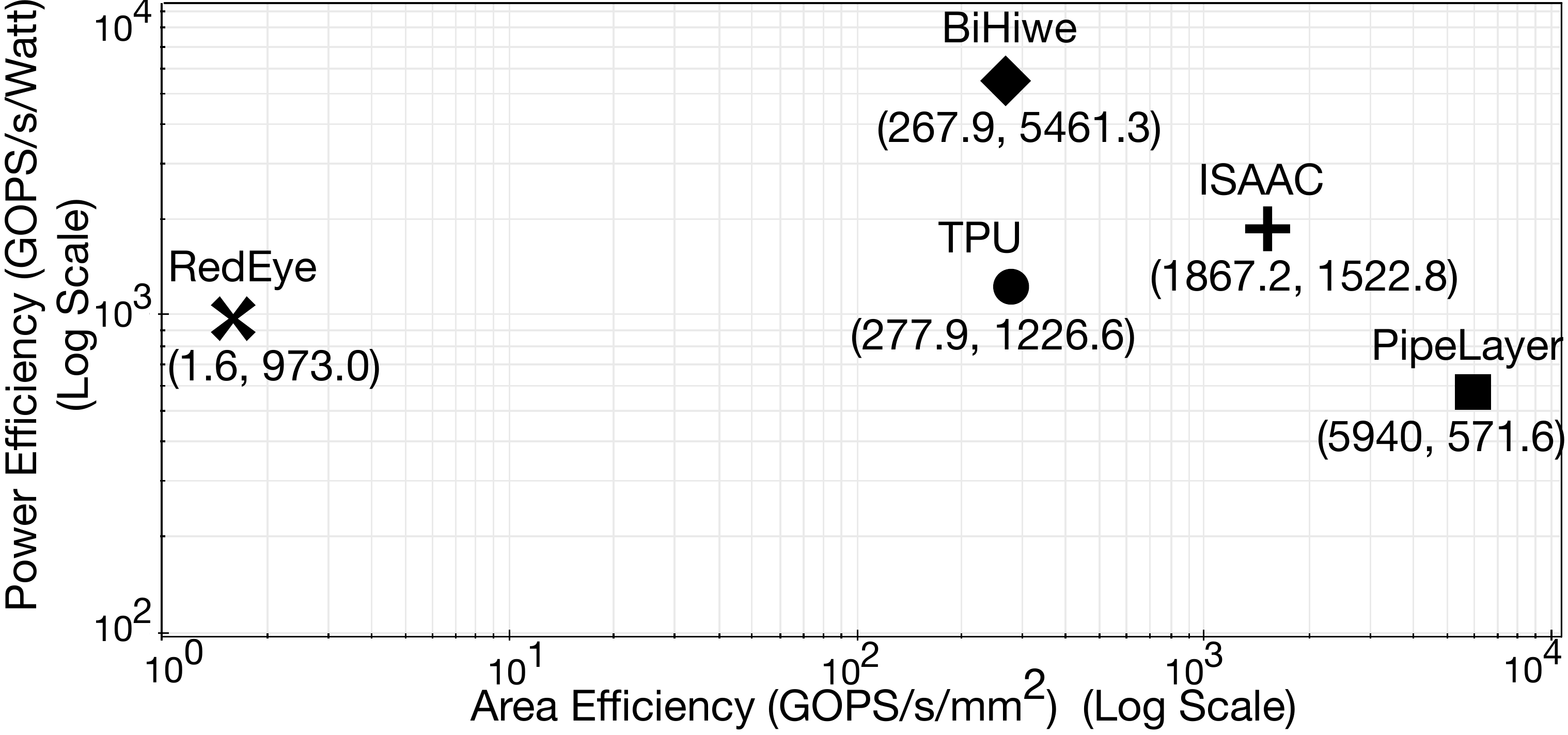}
	\vspace{1pt}
	\caption{Comparison with other accelerators.}
	\label{fig:others}
\vspace{-2.5ex}
\end{figure}

We also compare the power efficiency~(GOPS/s/Watt) and area efficiency~{GOPS/s/$mm^2$} of \bihiwe with other recent digital and analog accelerators.
Due to the lack of available raw performance/energy numbers for specific workloads, we use these metrics that is commensurate with comparisons for recent designs~\cite{yodann:arxiv:2017,  pipelayer:hpca:17, long2018reram}.
Figure~\ref{fig:others} depicts the peak power and area efficiency results.
On average for the evaluated benchmarks, \bihiwe achieves 72\% of its peak efficiency.
This information is not available in the publications for the other designs.

\niparagraph{Digital systolic: Google TPU~\cite{tpu:isca:2017}.} 
In comparison with TPU, which also uses systolic design, \atlass delivers 4.5$\times$ more peak power efficiency and almost the same area efficiency.
Leveraging the wide, interleaved, and bit-partitioned arithmetic with its switched-capacitor implementation in \atlass architecture, reduces the cost of MACC operations significantly compared with TPU which uses 8-bit digital logic, leading to significant improvement in power efficiency.

\niparagraph{Mixed-signal CMOS: RedEye~\cite{redeye:isca16}.}
RedEye is an in-sensor CNN accelerator baed on mixed-signal CMOS technology which also uses switched-capacitor circuitry for MACC operations.
Compared to RedEye, \atlass offers 5.5$\times$ better power efficiency and 167$\times$ better area efficiency.
Utilizing the proposed wide, interleaved, and bit-partitioned arithmetic amortizes the cost of ADC in \atlass by reducing its required resolution and sampling rate, leading to significant curtailment of ADC power and area, in contrast to RedEye.
%
%
%

\niparagraph{Analog Memristive designs~\cite{isaac:isca:2016, pipelayer:hpca:17}.}
Prior work in ISAAC~\cite{isaac:isca:2016} and PipeLayer~\cite{pipelayer:hpca:17} have explored analog memristive technology for DNN acceleration, which integrates both compute and storage within the same die, and offers higher compute density compared to traditional analog CMOS technology.
However, this increase in compute density comes at the cost of reduced power-efficiency.
Generally, memrisitive designs perform computations in the current domain, requiring the costly ADCs to sample the current-domain signals at the same rate as the compute/storage for memristors.
PipeLayer significantly reduces this cost.
Overall, compared to ISAAC and PipeLayer, \atlass improves the power efficiency by 3.6$\times$ and 9.6$\times$, respectively.
\subsubsection{Design Space Explorations}
\niparagraph{Design space exploration for bit-partitioning.}
\begin{figure}
	\centering
	\includegraphics[width=0.6\linewidth]{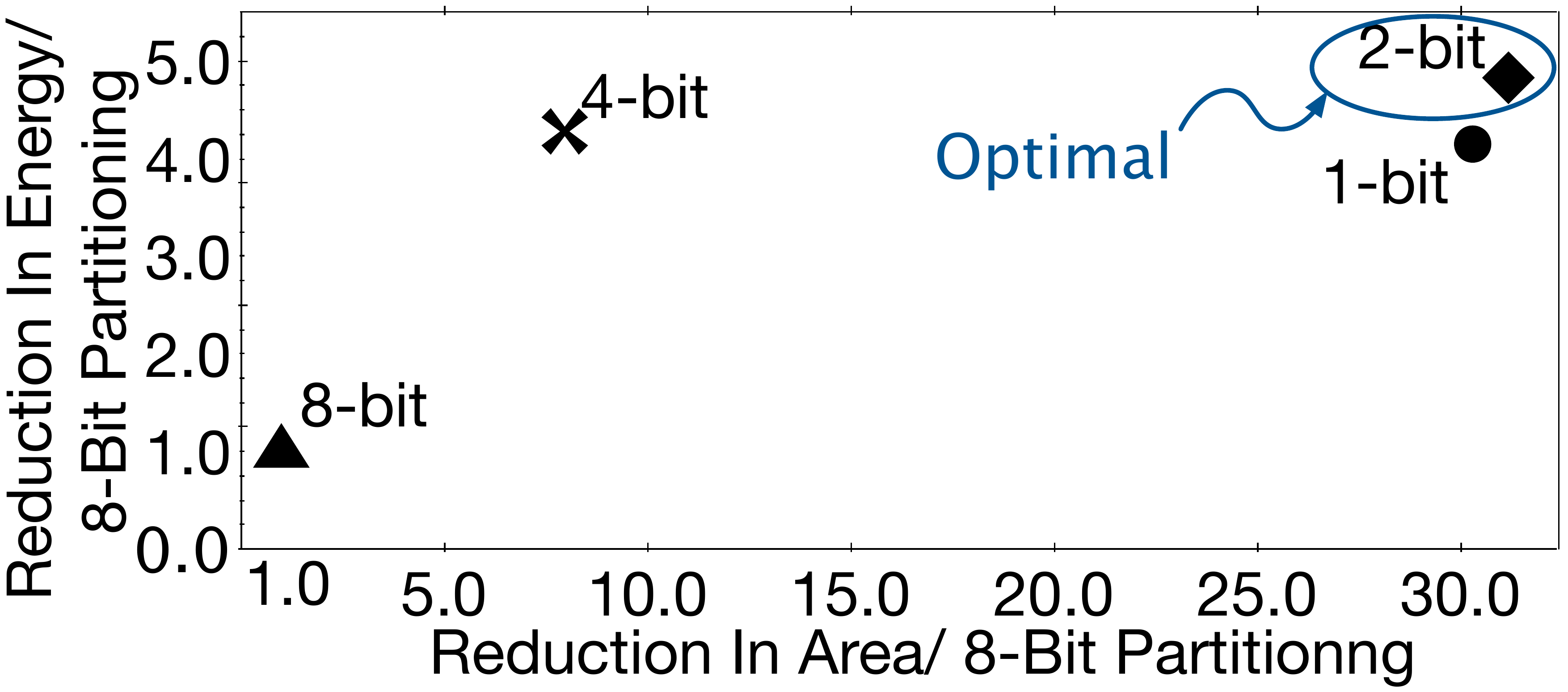}
	\vspace{1pt}
	\caption{Design space exploration for bit-partitioning.}
	\label{fig:bitp_design_space}
\vspace{-4ex}
\end{figure}
To evaluate the effectiveness of bit-partitioning, we perform a design space exploration with various bit-partitioned options.
Figure~\ref{fig:bitp_design_space} shows the reduction in energy and area compared to an 8-bit$\times$8-bit design when two vectors with 32 elements go under dot-product.
The other design points also perform 8-bit$\times$8-bit MACC operations while utilizing our wide and interleaved bit-partitioned arithmetic.
As depicted, the design with 2-bit partitioning strikes the best balance in energy and area with the switched-capacitor design of MACC units at 45~nm CMOS node.
The difference between 2-bit and 1-bit is that single-bit partitioning quadratically increases the number of low bitwidth MACCs from 16 (2-bit partitioning ) to 64 (1-bit partitioning) to support 8-bit operations.
This imposes disproportionate overhead that outweighs the benefit of decreasing each MACC units area and energy.

\niparagraph{Design space exploration for \msbpmacc configuration.}
\begin{figure}[!hb]
\vspace{-2.5ex}
	\centering
	\includegraphics[width=0.8\linewidth]{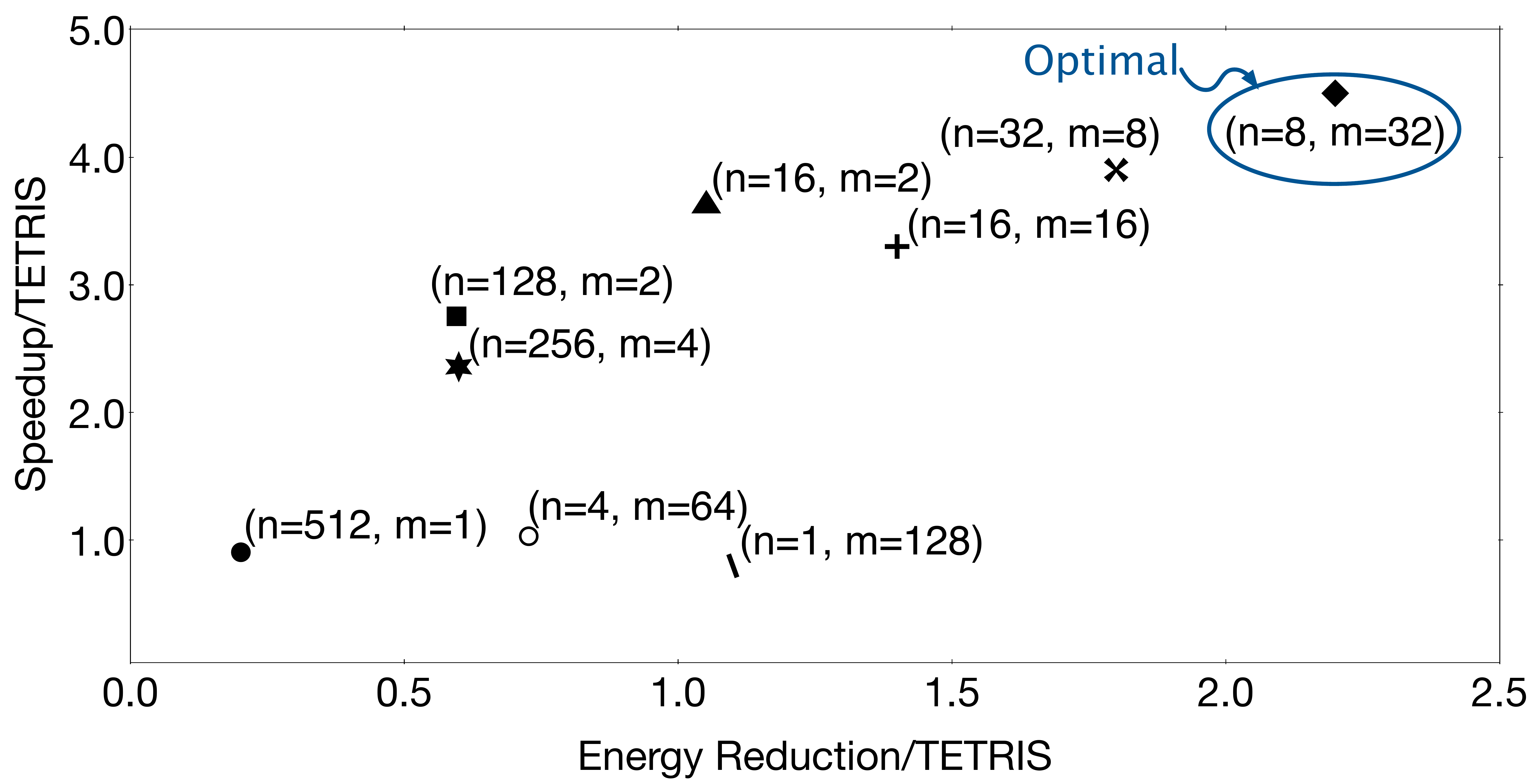}
	\vspace{1pt}
	\caption{Design space exploration for \msbpmacc.}
	\label{fig:msbpmacc_design_space}
\vspace{-2.8ex}
\end{figure}
The number of accumulation cycles ($m$) before the A/D conversion and the number of MACC units ($n$) are two main parameters of \msbpmacc which define resolution and the sample rate of the ADC, determining its power.
Figure~\ref{fig:msbpmacc_design_space} shows the design space exploration for different configurations of the \msbpmacc.
In a fixed power budget of $2$W for compute units, we measure the total runtime and energy of \atlass over the evaluated workloads which are normalized to \tetris.
As shown in Figure~\ref{fig:msbpmacc_design_space}, increasing number of MACCs, limits the number of accumulation cycles, consequently leading to using ADCs with high sample-rates.
Using high sample-rate ADCs significantly increases power, making the design less efficient.
On the other hand, increasing number of accumulation cycles, limits the number of MACCs, which restricts the number of \mswaggs that can be integrated into the design under the given power budget.
Overall, the optimal design point that delivers the best performance and energy is with eight MACC units and 32 accumulation cycles.

\niparagraph{Design space exploration for clustered architecture.}
\begin{figure}[]
\vspace{1ex}
	\centering
	\includegraphics[width=0.55\linewidth]{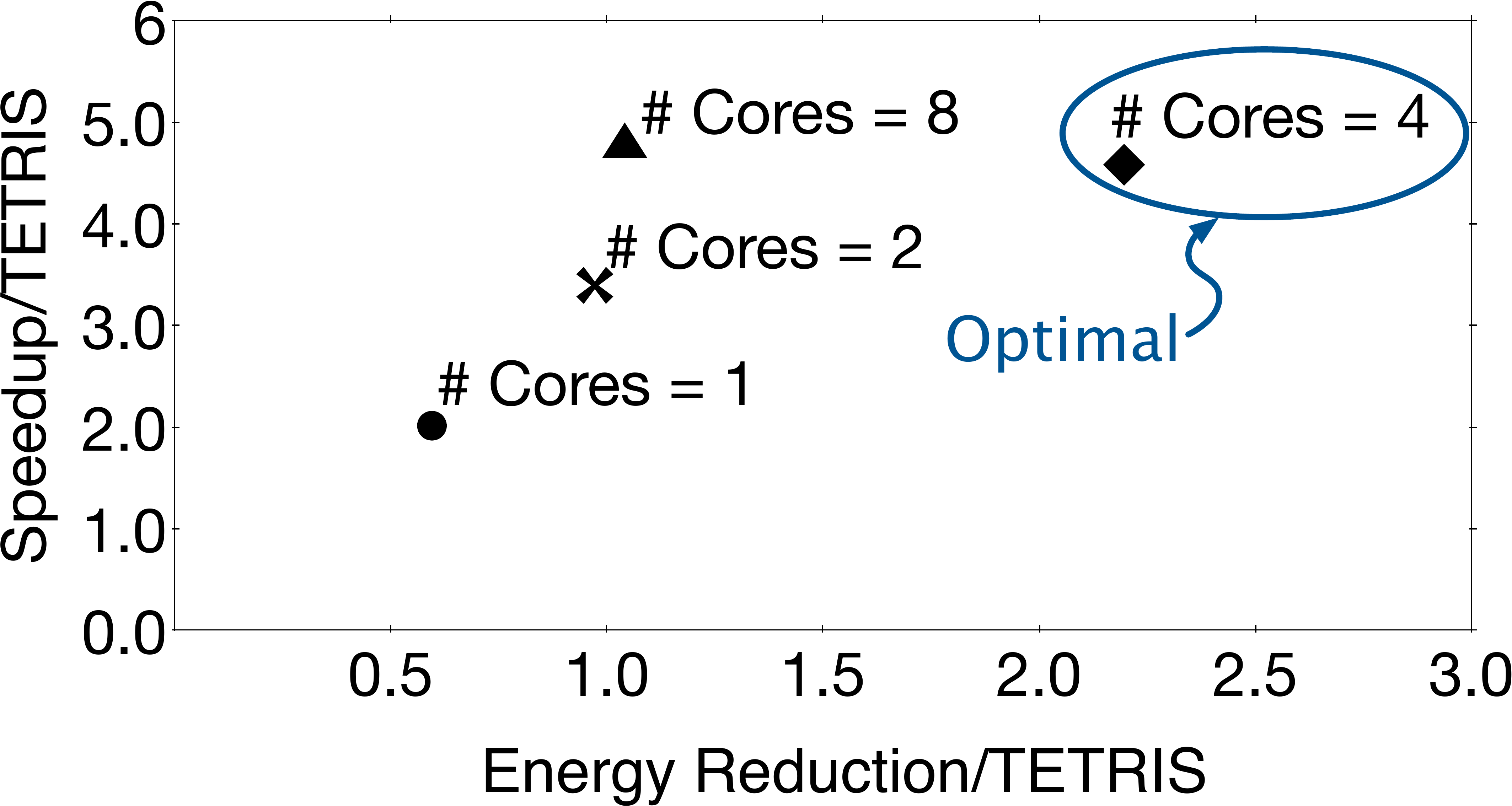}
	\caption{Design space exploration for \# core per cluster.}
	\label{fig:core-design-space}
	\vspace{-4ex}
\end{figure}
\atlass uses a hierarchical architecture with multiple cores in each vault.
Having a larger number of small cores for each vault yields increased utilization of compute resources, but requires data transfer across cores.
We explore the design space with 1, 2, 4, and 8 cores per cluster.%
As Figure\ref{fig:core-design-space} shows, \atlass with four cores per each vault (default configuration in \atlass) strikes the best balance between speedup and energy reduction.
Performance increases as we increase the number of cores per vault from 1 to 8.
However, the 8-core configuration results in a higher number of data accesses.
Therefore, the 4-core design point provides the optimal balance.

\subsubsection{Evaluation of Circuitry Non-Idealities}
%
%
Table~\ref{tab:retraining} shows the Top-1 accuracy with considering non-idealities, after fine-tuning, the ideal accuracy, and the final loss in accuracy.%
\begin{table}[!hb]
\centering
\vspace{-2ex}
\caption{Accuracy before and after fine-tuning.}
\label{tab:retraining}
\scalebox{0.45}{
\sf
\begin{tabular}{@{}cccccc@{}}
\toprule
\textbf{DNN Model} & \textbf{Dataset} & \textbf{\begin{tabular}[c]{@{}c@{}}Top-1 Accuracy\\ (With non-idealities)\end{tabular}} & \textbf{\begin{tabular}[c]{@{}c@{}}Top-1 Accuracy\\ (After fine-tuning)\end{tabular}} & \textbf{\begin{tabular}[c]{@{}c@{}}Top-1 Accuracy\\ (Ideal)\end{tabular}} & \textbf{\begin{tabular}[c]{@{}c@{}}Accuracy Loss\end{tabular}}\\ \midrule
\textbf{AlexNet} & Imagenet & 53.12\% & 56.64\% & 57.11\% & 0.47\%\\
\textbf{CIFAR-10} & CIFAR-10 & 90.82\% & 91.01\% & 91.03\% & 0.02\%\\
\textbf{GoogLeNet} & Imagenet & 67.15\% & 68.39\% & 68.72\% & 0.33\%\\
\textbf{ResNet-18} & Imagenet & 66.91\% & 68.96\% & 68.98\% & 0.02\%\\
\textbf{ResNet-50} & Imagenet & 74.5\% & 75.21\% & 75.25\% & 0.04\%\\
\textbf{VGG-16} & Imagenet & 70.31\% & 71.28\% & 71.46\% & 0.18\%\\
\textbf{VGG-19} & Imagenet & 73.24\% & 74.20\% & 74.52\% & 0.32\%\\
\textbf{YOLOv3} & Imagenet & 75.92\% & 77.1\% & 77.22\% & 0.21\%\\
	\
\textbf{PTB-RNN} & Penn TreeBank & 1.1 BPC & 1.6 BPC & 1.1 BPC & 0.0 BPC\\
\textbf{PTB-LSTM} & Penn TreeBank & 97 PPW & 170 PPW & 97 PPW & 0.0 PPW\\ \bottomrule
\end{tabular}
}
\vspace{-2.5ex}
\end{table}
As shown in Table~\ref{tab:retraining}, some of the networks, namely \bench{AlexNet} and \bench{ResNet-18}, are more sensitive to the non-idealities, leading to a higher initial accuracy degradation.
To recover the accuracy loss due to the circuitry non-idealities, we perform a fine-tuning step for a few epochs.
By performing this fine-tuning step, the accuracy loss of the \bench{CIFAR-10}, \bench{ResNet-18}, and \bench{ResNet-50} networks is fully recovered~(loss is less than 0.04\%) which within these networks, \bench{CIFAR-10} and \bench{ResNet-50} are more robust to non-idealities.
%
The accuracy loss for other networks is below 0.5\% which within those \bench{AlexNet} has the maximum loss.
The final two networks, namely \bench{PTB-RNN} and \bench{PTB-LSTM} perform character-level and word-level language modeling, respectively.
The accuracy for these two networks is measured in Bits-Per-Character (BPC) and Perplexity-per-Word (PPW), respectively.
Both \bench{PTB-RNN} and \bench{PTB-LSTM} recover all the loss after fine-tuning.
The final results after fine-tuning step show the effectiveness of this approach in recovering the accuracy loss due to the non-idealities pertinent to analog computation.

\vspace{-1ex}
\section{Related Work}
\label{sec:related}
\vspace{-0.5ex}
%
%
There is a large body of work on digital accelerators for DNNs~\cite{npu:cacm,dadiannao:micro:2014,tetris:asplos:2017, tartan:arxiv:2017, tabla:hpca:2016, cambricon-x:micro:2016,cnvlutin:isca:2016,stripes:micro:2016,dnnweaver:micro:2016, brainwave:hotchips:2017, scnn:isca:2017,yodann:arxiv:2017, eie:isca:2016, eyeriss:isca:2016, eyeriss:jssc:2017, neurocube:isca:16, tpu:isca:2017,diannao:asplos:2014, bitfusion:isca18, snapea:isca:2018, ucnn:isca18, unpu:isscc:2018}.
Mixed-signal acceleration has also been explored previously for neural network~\cite{tsividis1987switched, anpu} and is gaining traction again for the deep models~\cite{isaac:isca:2016, promise:isca18, redeye:isca16, Mour:charge, switched-capacitor, Mour:cifar10, lca:vlsi2017,  zhang201518, passive:switch}.
This paper fundamentally differs from these inspiring efforts as it delves into the mathematics of basic operations in DNNs, reformulates and defines the wide, interleaved, and bit-partitioned approach to overcome the challenges of mixed-signal acceleration.
By partitioning and re-aggregating the low-bitwidth MACC operations, this paper addresses the limited range of encoding and reduces the cost of cross-domain conversions.
Additionally, it combines the proposed mathematical reformulation with switched-capacitor circuitry to share and delay A/D conversions, which amortizes their cost and reduce their rate, respectively.
Below, we discuss the most related works.

\niparagraph{Switched-capacitor design.}
Switched-capacitor circuits~\cite{gray2001analysis} have a long history, having been mainly used for designing amplifiers\cite{scap:amp}, A/D and D/A converters\cite{scap:converter} and filters\cite{scap:filter}.
Similar to resistive circuits, they have been used even for the previous generation of neural networks~\cite{tsividis1987switched}.
More recently, they have also been used for matrix multiplication\cite{Mour:SCdotproduct, passive:switch}, which can benefit DNNs.
This work takes inspiration from these efforts but differes from them in that it defines and leverages wide, interleaved, and bit-partitioned reformulation of DNN operations.
Additionally, it offers a comprehensive architecture that can accelerate a wide variety of DNNs.

\niparagraph{Programmable mixed-signal accelerators.}
PROMISE~\cite{promise:isca18} offers a mixed-signal architecture that integrates analog units within the SRAM memory blocks.
RedEye\cite{redeye:isca16} is a low-power near-sensor mixed-signal accelerator that uses charge-domain computations.
These works do not offer wide interleavings of bit-partitioned basic operations as described in this paper.

%
\niparagraph{Fixed-functional mixed-signal accelerators.}
They are designed for a specific DNN. 
Some focus on handwritten digit classification~\cite{Mour:SCdotproduct, timedomain:jssc2017} or binarized mixed-signal acceleration of CIFAR-10 images~\cite{Mour:cifar10}.
Another work focuses on spiking neural networks' acceleration~\cite{lca:vlsi2017}.
In contrast, our design is programmable and supports interleaved bit-partitioning.
  
%
\niparagraph{Resistive memory accelerators.}
There is a large body of work using resistive memory~\cite{isaac:isca:2016, pipelayer:hpca:17, long2018reram, qiao2018atomlayer, ji2018recom, li2018reram,  chen2017accelerator, chi2016prime}.
We provided a direct comparison to ISAAC~\cite{isaac:isca:2016} and PipeLayer~\cite{pipelayer:hpca:17}.
ISAAC~\cite{isaac:isca:2016} most notably introduces the concept of temporally bit-serial operations, also explored in PRIME~\cite{prime:isca:2016}, and is augmented with the concept of spike-base data scheme in PipeLayer~\cite{pipelayer:hpca:17}. 
\bihiwe, in contrast, formulates a partitioning that spatially groups lower-bitwidth MACCs across different vector elements and performs them in-parallel.
%
%
PRIME does not provide absolute measurements and its simulated baseline is not available for a head-to-head comparison.
PRIME also uses multiple truncations that change the mathematics.
Conversely, our formulation does not induce truncation or mathematical changes.

\vspace{-1ex}
\section{Conclusion}
\label{sec:conclusion}
\vspace{-0.5ex}

%

This work proposes wide, interleaved, and bit-partitioned arithmetic to overcome two key challenges in mixed-signal acceleration of DNNs: limited encoding range, and costly A/D conversions.
This bit-partitioned arithmetic enables rearranging the highly parallel MACC operations in modern DNNs into wide low-bitwidth computations that are mapped efficiently to mixed-signal units.
Further, these units operate in charge domain using switched-capacitor circuitry and reduce the rate of A/D conversions by accumulating partial results in the charge domain.
The resulting microarchitecture, named \atlass, offers significant  benefits over its state-of-the-art analog and digital counterparts.
These encouraging results suggest that the combination of mathematical insights with architectural innovations can enable new avenues in DNN acceleration.
\vspace{-1ex}


\bibliography{paper.bib}

\begin{thebibliography}{88}
\providecommand{\natexlab}[1]{#1}
\providecommand{\url}[1]{\texttt{#1}}
\expandafter\ifx\csname urlstyle\endcsname\relax
  \providecommand{\doi}[1]{doi: #1}\else
  \providecommand{\doi}{doi: \begingroup \urlstyle{rm}\Url}\fi

\bibitem[{Niehues} et~al.(2018){Niehues}, {Pham}, {Ha}, {Sperber}, and
  {Waibel}]{dnnspeech}
J.~{Niehues}, N.-Q. {Pham}, T.-L. {Ha}, M.~{Sperber}, and A.~{Waibel}.
\newblock {Low-Latency Neural Speech Translation}.
\newblock \emph{ArXiv e-prints}, August 2018.

\bibitem[{Mo} and {Sattar}(2018)]{dnndriving}
J.~{Mo} and J.~{Sattar}.
\newblock {SafeDrive: Enhancing Lane Appearance for Autonomous and Assisted
  Driving Under Limited Visibility}.
\newblock \emph{ArXiv e-prints}, July 2018.

\bibitem[{Li} et~al.(2018){Li}, {Shu}, {Su}, {Feng}, and {Wang}]{dnnsearch}
R.~{Li}, Y.~{Shu}, J.~{Su}, H.~{Feng}, and J.~{Wang}.
\newblock {Using deep Residual Network to search for
  galaxy-Ly$\{$$\backslash$alpha$\}$ emitter lens candidates based on
  spectroscopic-selection}.
\newblock \emph{ArXiv e-prints}, July 2018.

\bibitem[{Rohde} et~al.(2018){Rohde}, {Bonner}, {Dunlop}, {Vasile}, and
  {Karatzoglou}]{dnnadvertising}
D.~{Rohde}, S.~{Bonner}, T.~{Dunlop}, F.~{Vasile}, and A.~{Karatzoglou}.
\newblock {RecoGym: A Reinforcement Learning Environment for the problem of
  Product Recommendation in Online Advertising}.
\newblock \emph{ArXiv e-prints}, August 2018.

\bibitem[{Grabec} et~al.(2018){Grabec}, {{\v S}vegl}, and {Sok}]{dnnmedical}
I.~{Grabec}, E.~{{\v S}vegl}, and M.~{Sok}.
\newblock {Development of a sensory-neural network for medical diagnosing}.
\newblock \emph{ArXiv e-prints}, July 2018.

\bibitem[Esmaeilzadeh et~al.(2011)Esmaeilzadeh, Blem, St.~Amant, Sankaralingam,
  and Burger]{dark_silicon:isca}
Hadi Esmaeilzadeh, Emily Blem, Renee St.~Amant, Karthikeyan Sankaralingam, and
  Doug Burger.
\newblock Dark silicon and the end of multicore scaling.
\newblock In \emph{ISCA}, 2011.

\bibitem[Hardavellas et~al.(2011)Hardavellas, Ferdman, Falsafi, and
  Ailamaki]{dark_silicon:babak}
N.~Hardavellas, M.~Ferdman, B.~Falsafi, and A.~Ailamaki.
\newblock Toward dark silicon in servers.
\newblock \emph{IEEE Micro}, 31\penalty0 (4):\penalty0 6--15, July--Aug. 2011.

\bibitem[Venkatesh et~al.(2010)Venkatesh, Sampson, Goulding, Garcia, Bryksin,
  Lugo-Martinez, Swanson, and Taylor]{ccores}
Ganesh Venkatesh, Jack Sampson, Nathan Goulding, Saturnino Garcia, Vladyslav
  Bryksin, Jose Lugo-Martinez, Steven Swanson, and Michael~Bedford Taylor.
\newblock Conservation cores: Reducing the energy of mature computations.
\newblock In \emph{ASPLOS}, 2010.

\bibitem[Zhang et~al.(2015{\natexlab{a}})Zhang, Li, Sun, Guan, Xiao, and
  Cong]{dnnoptimizing:fpga:2015}
Chen Zhang, Peng Li, Guangyu Sun, Yijin Guan, Bingjun Xiao, and Jason Cong.
\newblock Optimizing fpga-based accelerator design for deep convolutional
  neural networks.
\newblock In \emph{FPGA}, 2015{\natexlab{a}}.

\bibitem[Esmaeilzadeh et~al.(2013)Esmaeilzadeh, Sampson, Ceze, and
  Burger]{npu:cacm}
Hadi Esmaeilzadeh, Adrian Sampson, Luis Ceze, and Doug Burger.
\newblock Neural acceleration for general-purpose approximate programs.
\newblock \emph{to apear in Commun. ACM}, 2013.

\bibitem[Chen et~al.(2014{\natexlab{a}})Chen, Luo, Liu, Zhang, He, Wang, Li,
  Chen, Xu, Sun, et~al.]{dadiannao:micro:2014}
Yunji Chen, Tao Luo, Shaoli Liu, Shijin Zhang, Liqiang He, Jia Wang, Ling Li,
  Tianshi Chen, Zhiwei Xu, Ninghui Sun, et~al.
\newblock Dadiannao: A machine-learning supercomputer.
\newblock In \emph{MICRO}, 2014{\natexlab{a}}.

\bibitem[Gao et~al.(2017{\natexlab{a}})Gao, Pu, Yang, Horowitz, and
  Kozyrakis]{tetris:asplos:2017}
Mingyu Gao, Jing Pu, Xuan Yang, Mark Horowitz, and Christos Kozyrakis.
\newblock Tetris: Scalable and efficient neural network acceleration with 3d
  memory.
\newblock In \emph{ASPLOS}, 2017{\natexlab{a}}.

\bibitem[Delmas et~al.(2017)Delmas, Sharify, Judd, and
  Moshovos]{tartan:arxiv:2017}
Alberto Delmas, Sayeh Sharify, Patrick Judd, and Andreas Moshovos.
\newblock Tartan: Accelerating fully-connected and convolutional layers in deep
  learning networks by exploiting numerical precision variability.
\newblock \emph{arXiv}, 2017.

\bibitem[Mahajan et~al.(2016)Mahajan, Park, Amaro, Sharma, Yazdanbakhsh, Kim,
  and Esmaeilzadeh]{tabla:hpca:2016}
Divya Mahajan, Jongse Park, Emmanuel Amaro, Hardik Sharma, Amir Yazdanbakhsh,
  Joon Kim, and Hadi Esmaeilzadeh.
\newblock {TABLA}: A unified template-based framework for accelerating
  statistical machine learning.
\newblock In \emph{HPCA}, 2016.

\bibitem[Zhang et~al.(2016)Zhang, Du, Zhang, Lan, Liu, Li, Guo, Chen, and
  Chen]{cambricon-x:micro:2016}
Shijin Zhang, Zidong Du, Lei Zhang, Huiying Lan, Shaoli Liu, Ling Li, Qi~Guo,
  Tianshi Chen, and Yunji Chen.
\newblock Cambricon-x: An accelerator for sparse neural networks.
\newblock In \emph{MICRO}, 2016.

\bibitem[Albericio et~al.(2016)Albericio, Judd, Hetherington, Aamodt, Jerger,
  and Moshovos]{cnvlutin:isca:2016}
Jorge Albericio, Patrick Judd, Tayler Hetherington, Tor Aamodt, Natalie~Enright
  Jerger, and Andreas Moshovos.
\newblock Cnvlutin: ineffectual-neuron-free deep neural network computing.
\newblock In \emph{ISCA}, 2016.

\bibitem[Judd et~al.(2016)Judd, Albericio, Hetherington, Aamodt, and
  Moshovos]{stripes:micro:2016}
Patrick Judd, Jorge Albericio, Tayler Hetherington, Tor~M Aamodt, and Andreas
  Moshovos.
\newblock Stripes: Bit-serial deep neural network computing.
\newblock In \emph{MICRO}, 2016.

\bibitem[Sharma et~al.(2016)Sharma, Park, Mahajan, Amaro, Kim, Shao, Misra, and
  Esmaeilzadeh]{dnnweaver:micro:2016}
Hardik Sharma, Jongse Park, Divya Mahajan, Emmanuel Amaro, Joon Kim, Chenkai
  Shao, Asit Misra, and Hadi Esmaeilzadeh.
\newblock From high-level deep neural models to fpgas.
\newblock In \emph{MICRO}, 2016.

\bibitem[Chung et~al.(2017)Chung, Fowers, Ovtcharov, Papamichael, Caulfield,
  Massengil, Liu, Lo, Alkalay, Haselman, Boehn, Firestein, Forin, Gatlin,
  Ghandi, Heil, Holohan, Juhasz, Kovvuri, Lanka, van Megen, Mukhortov, Patel,
  Reinhardt, Sapek, Seera, Sridharan, Woods, Yi-Xiao, Zhao, and
  Burger]{brainwave:hotchips:2017}
Eric Chung, Jeremy Fowers, Kalin Ovtcharov, Michael Papamichael, Adrian
  Caulfield, Todd Massengil, Ming Liu, Daniel Lo, Shlomi Alkalay, Michael
  Haselman, Christian Boehn, Oren Firestein, Alessandro Forin, Kang~Su Gatlin,
  Mahdi Ghandi, Stephen Heil, Kyle Holohan, Tamas Juhasz, Ratna~Kumar Kovvuri,
  Sitaram Lanka, Friedel van Megen, Dima Mukhortov, Prerak Patel, Steve
  Reinhardt, Adam Sapek, Raja Seera, Balaji Sridharan, Lisa Woods, Phillip
  Yi-Xiao, Ritchie Zhao, and Doug Burger.
\newblock Accelerating persistent neural networks at datacenter scale.
\newblock In \emph{HotChips}, 2017.

\bibitem[Parashar et~al.(2017)Parashar, Rhu, Mukkara, Puglielli, Venkatesan,
  Khailany, Emer, Keckler, and Dally]{scnn:isca:2017}
Angshuman Parashar, Minsoo Rhu, Anurag Mukkara, Antonio Puglielli, Rangharajan
  Venkatesan, Brucek Khailany, Joel Emer, Stephen~W Keckler, and William~J
  Dally.
\newblock {SCNN: An Accelerator for Compressed-sparse Convolutional Neural
  Networks}.
\newblock In \emph{ISCA}, 2017.

\bibitem[Andri et~al.(2016)Andri, Cavigelli, Rossi, and
  Benini]{yodann:arxiv:2017}
Renzo Andri, Lukas Cavigelli, Davide Rossi, and Luca Benini.
\newblock Yodann: An ultra-low power convolutional neural network accelerator
  based on binary weights.
\newblock \emph{arXiv}, 2016.

\bibitem[Han et~al.(2016)Han, Liu, Mao, Pu, Pedram, Horowitz, and
  Dally]{eie:isca:2016}
Song Han, Xingyu Liu, Huizi Mao, Jing Pu, Ardavan Pedram, Mark~A Horowitz, and
  William~J Dally.
\newblock Eie: efficient inference engine on compressed deep neural network.
\newblock In \emph{ISCA}, 2016.

\bibitem[Chen et~al.(2016)Chen, Emer, and Sze]{eyeriss:isca:2016}
Yu-Hsin Chen, Joel Emer, and Vivienne Sze.
\newblock Eyeriss: A spatial architecture for energy-efficient dataflow for
  convolutional neural networks.
\newblock In \emph{ISCA}, 2016.

\bibitem[Chen et~al.(2017{\natexlab{a}})Chen, Krishna, Emer, and
  Sze]{eyeriss:jssc:2017}
Yu-Hsin Chen, Tushar Krishna, Joel~S Emer, and Vivienne Sze.
\newblock Eyeriss: An energy-efficient reconfigurable accelerator for deep
  convolutional neural networks.
\newblock \emph{JSSC}, 2017{\natexlab{a}}.

\bibitem[Kim et~al.(2016)Kim, Kung, Chai, Yalamanchili, and
  Mukhopadhyay]{neurocube:isca:16}
Duckhwan Kim, Jaeha Kung, Sek Chai, Sudhakar Yalamanchili, and Saibal
  Mukhopadhyay.
\newblock Neurocube: A programmable digital neuromorphic architecture with
  high-density 3d memory.
\newblock In \emph{Computer Architecture (ISCA), 2016 ACM/IEEE 43rd Annual
  International Symposium on}, pages 380--392. IEEE, 2016.

\bibitem[Jouppi et~al.(2017)Jouppi, Young, Patil, Patterson, Agrawal, Bajwa,
  Bates, Bhatia, Boden, Borchers, et~al.]{tpu:isca:2017}
Norman~P Jouppi, Cliff Young, Nishant Patil, David Patterson, Gaurav Agrawal,
  Raminder Bajwa, Sarah Bates, Suresh Bhatia, Nan Boden, Al~Borchers, et~al.
\newblock In-datacenter performance analysis of a tensor processing unit.
\newblock In \emph{ISCA}, 2017.

\bibitem[Chen et~al.(2014{\natexlab{b}})Chen, Du, Sun, Wang, Wu, Chen, and
  Temam]{diannao:asplos:2014}
Tianshi Chen, Zidong Du, Ninghui Sun, Jia Wang, Chengyong Wu, Yunji Chen, and
  Olivier Temam.
\newblock Diannao: a small-footprint high-throughput accelerator for ubiquitous
  machine-learning.
\newblock In \emph{ASPLOS}, 2014{\natexlab{b}}.

\bibitem[Sharma et~al.()Sharma, Park, Suda, Lai, Chau, Chandra, and
  Esmaeilzadeh]{bitfusion:isca18}
Hardik Sharma, Jongse Park, Naveen Suda, Liangzhen Lai, Benson Chau, Vikas
  Chandra, and Hadi Esmaeilzadeh.
\newblock Bit fusion: Bit-level dynamically composable architecture for
  accelerating deep neural networks.

\bibitem[Aklaghi et~al.(2018)Aklaghi, Yazdanbakhsh, Samadi, Esmaeilzadeh, and
  K.~Gupta]{snapea:isca:2018}
Vahide Aklaghi, Amir Yazdanbakhsh, Kambiz Samadi, Hadi Esmaeilzadeh, and Rajesh
  K.~Gupta.
\newblock Snapea: Predictive early activation for reducing computation in deep
  convolutional neural networks.
\newblock In \emph{ISCA}, 2018.

\bibitem[Hegde et~al.(2018)Hegde, Yu, Agrawal, Yan, Pellauer, and
  Fletcher]{ucnn:isca18}
Kartik Hegde, Jiyong Yu, Rohit Agrawal, Mengjia Yan, Michael Pellauer, and
  Christopher~W Fletcher.
\newblock Ucnn: Exploiting computational reuse in deep neural networks via
  weight repetition.
\newblock \emph{arXiv preprint arXiv:1804.06508}, 2018.

\bibitem[Lee et~al.(2018)Lee, Kim, Kang, Shin, Kim, and Yoo]{unpu:isscc:2018}
Jinmook Lee, Changhyeon Kim, Sanghoon Kang, Dongjoo Shin, Sangyeob Kim, and
  Hoi-Jun Yoo.
\newblock Unpu: A 50.6 tops/w unified deep neural network accelerator with
  1b-to-16b fully-variable weight bit-precision.
\newblock In \emph{ISSCC}, 2018.

\bibitem[Shafiee et~al.(2016)Shafiee, Nag, Muralimanohar, Balasubramonian,
  Strachan, Hu, Williams, and Srikumar]{isaac:isca:2016}
Ali Shafiee, Anirban Nag, Naveen Muralimanohar, Rajeev Balasubramonian,
  John~Paul Strachan, Miao Hu, R~Stanley Williams, and Vivek Srikumar.
\newblock Isaac: A convolutional neural network accelerator with in-situ analog
  arithmetic in crossbars.
\newblock In \emph{ISCA}, 2016.

\bibitem[Srivastava et~al.(2018)Srivastava, Kang, Gonugondla, Lim, Choi, Adve,
  Kim, and Shanbhag]{promise:isca18}
Prakalp Srivastava, Mingu Kang, Sujan~K Gonugondla, Sungmin Lim, Jungwook Choi,
  Vikram Adve, Nam~Sung Kim, and Naresh Shanbhag.
\newblock Promise: An end-to-end design of a programmable mixed-signal
  accelerator for machine-learning algorithms.
\newblock In \emph{2018 ACM/IEEE 45th Annual International Symposium on
  Computer Architecture (ISCA)}. IEEE, 2018.

\bibitem[Tsividis and Anastassiou(1987)]{tsividis1987switched}
YP~Tsividis and D~Anastassiou.
\newblock Switched-capacitor neural networks.
\newblock \emph{Electronics Letters}, 23\penalty0 (18):\penalty0 958--959,
  1987.

\bibitem[LiKamWa et~al.(2016)LiKamWa, Hou, Gao, Polansky, and
  Zhong]{redeye:isca16}
Robert LiKamWa, Yunhui Hou, Julian Gao, Mia Polansky, and Lin Zhong.
\newblock Redeye: analog convnet image sensor architecture for continuous
  mobile vision.
\newblock In \emph{ACM SIGARCH Computer Architecture News}, volume~44, pages
  255--266. IEEE Press, 2016.

\bibitem[Bankman and Murmann(2015)]{Mour:charge}
Daniel Bankman and Boris Murmann.
\newblock Passive charge redistribution digital-to-analogue multiplier.
\newblock \emph{Electronics Letters}, 51\penalty0 (5):\penalty0 386--388, 2015.

\bibitem[Lee and Wong(2017{\natexlab{a}})]{switched-capacitor}
E.~H. Lee and S.~S. Wong.
\newblock Analysis and design of a passive switched-capacitor matrix multiplier
  for approximate computing.
\newblock \emph{IEEE Journal of Solid-State Circuits}, 52\penalty0
  (1):\penalty0 261--271, Jan 2017{\natexlab{a}}.
\newblock ISSN 0018-9200.
\newblock \doi{10.1109/JSSC.2016.2599536}.

\bibitem[Bankman et~al.(2018)Bankman, Yang, Moons, Verhelst, and
  Murmann]{Mour:cifar10}
Daniel Bankman, Lita Yang, Bert Moons, Marian Verhelst, and Boris Murmann.
\newblock An always-on 3.8 $\mu$j/86\% cifar-10 mixed-signal binary cnn
  processor with all memory on chip in 28nm cmos.
\newblock In \emph{Solid-State Circuits Conference-(ISSCC), 2018 IEEE
  International}, pages 222--224. IEEE, 2018.

\bibitem[Buhler et~al.(2017)Buhler, Brown, Li, Chen, Zhang, and
  Flynn]{lca:vlsi2017}
Fred~N Buhler, Peter Brown, Jiabo Li, Thomas Chen, Zhengya Zhang, and Michael~P
  Flynn.
\newblock A 3.43 tops/w 48.9 pj/pixel 50.1 nj/classification 512 analog neuron
  sparse coding neural network with on-chip learning and classification in 40nm
  cmos.
\newblock In \emph{VLSI Circuits, 2017 Symposium on}, pages C30--C31. IEEE,
  2017.

\bibitem[St.~Amant et~al.(2014)St.~Amant, Yazdanbakhsh, Park, Thwaites,
  Esmaeilzadeh, Hassibi, Ceze, and Burger]{anpu}
Ren{\'e}e St.~Amant, Amir Yazdanbakhsh, Jongse Park, Bradley Thwaites, Hadi
  Esmaeilzadeh, Arjang Hassibi, Luis Ceze, and Doug Burger.
\newblock General-purpose code acceleration with limited-precision analog
  computation.
\newblock In \emph{ISCA}, 2014.

\bibitem[Zhang et~al.(2015{\natexlab{b}})Zhang, Wang, and Verma]{zhang201518}
Jintao Zhang, Zhuo Wang, and Naveen Verma.
\newblock 18.4 a matrix-multiplying adc implementing a machine-learning
  classifier directly with data conversion.
\newblock In \emph{Solid-State Circuits Conference-(ISSCC), 2015 IEEE
  International}, pages 1--3. IEEE, 2015{\natexlab{b}}.

\bibitem[Lee and Wong(2017{\natexlab{b}})]{passive:switch}
Edward~H Lee and S~Simon Wong.
\newblock {Analysis and Design of a Passive Switched-Capacitor Matrix
  Multiplier for Approximate Computing}.
\newblock \emph{IEEE Journal of Solid-State Circuits}, 52\penalty0
  (1):\penalty0 261--271, 2017{\natexlab{b}}.

\bibitem[Gray et~al.(2001)Gray, Hurst, Meyer, and Lewis]{gray2001analysis}
Paul~R Gray, Paul Hurst, Robert~G Meyer, and Stephen Lewis.
\newblock \emph{Analysis and design of analog integrated circuits}.
\newblock Wiley, 2001.

\bibitem[Chi et~al.(2016{\natexlab{a}})Chi, Li, Xu, Zhang, Zhao, Liu, Wang, and
  Xie]{prime:isca:2016}
Ping Chi, Shuangchen Li, Cong Xu, Tao Zhang, Jishen Zhao, Yongpan Liu, Yu~Wang,
  and Yuan Xie.
\newblock Prime: A novel processing-in-memory architecture for neural network
  computation in reram-based main memory.
\newblock In \emph{ISCA}, 2016{\natexlab{a}}.

\bibitem[Sharify et~al.(2017)Sharify, Lascorz, Judd, and
  Moshovos]{loom:arxiv:2017}
Sayeh Sharify, Alberto~Delmas Lascorz, Patrick Judd, and Andreas Moshovos.
\newblock Loom: Exploiting weight and activation precisions to accelerate
  convolutional neural networks.
\newblock \emph{arXiv}, 2017.

\bibitem[Gao et~al.(2017{\natexlab{b}})Gao, Pu, Yang, Horowitz, and
  Kozyrakis]{tetris:simulator}
Mingyu Gao, Jing Pu, Xuan Yang, Mark Horowitz, and Christos Kozyrakis.
\newblock Tetris: Scalable and efficient neural network acceleration with 3d
  memory.
\newblock \url{https://github.com/stanford-mast/nn_dataflow},
  2017{\natexlab{b}}.

\bibitem[Li and Pedram(2017)]{li2017caterpillar}
Yuanfang Li and Ardavan Pedram.
\newblock Caterpillar: Coarse grain reconfigurable architecture for
  accelerating the training of deep neural networks.
\newblock In \emph{Application-specific Systems, Architectures and Processors
  (ASAP), 2017 IEEE 28th International Conference on}, pages 1--10. IEEE, 2017.

\bibitem[Upadhyay and Roy~Chowdhury(2015)]{8-bit-mult}
Himani Upadhyay and Shubhajit Roy~Chowdhury.
\newblock A high speed and low power 8 bit x 8 bit multiplier design using
  novel two transistor (2t) xor gates.
\newblock \emph{Journal of Low Power Electronics}, 01 2015.
\newblock \doi{10.1166/jolpe.2015.1362}.

\bibitem[Consortium et~al.(2013)]{HMC:spec}
Hybrid Memory~Cube Consortium et~al.
\newblock Hybrid memory cube specification 1.0.
\newblock \emph{Last Revision Jan}, 2013.

\bibitem[Jeddeloh and Keeth(2012)]{HMC:vlsi}
Joe Jeddeloh and Brent Keeth.
\newblock Hybrid memory cube new dram architecture increases density and
  performance.
\newblock In \emph{VLSI Technology (VLSIT), 2012 Symposium on}, pages 87--88.
  IEEE, 2012.

\bibitem[Yazdanbakhsh et~al.(2018)Yazdanbakhsh, Falahati, Wolfe, Samadi,
  Esmaeilzadeh, and Kim]{ganax:isca:2018}
Amir Yazdanbakhsh, Hajar Falahati, Philip~J. Wolfe, Kambiz Samadi, Hadi
  Esmaeilzadeh, and Nam~Sung Kim.
\newblock {GANAX: A Unified SIMD-MIMD Acceleration for Generative Adversarial
  Network}.
\newblock In \emph{ISCA}, 2018.

\bibitem[Ismail and Fiez(1994)]{analog:layout}
Mohammed Ismail and Terri Fiez.
\newblock \emph{Analog VLSI: signal and information processing}, volume 166.
\newblock McGraw-Hill New York, 1994.

\bibitem[Tripathi and Murmann(2014)]{fringe:cap}
Vaibhav Tripathi and Boris Murmann.
\newblock Mismatch characterization of small metal fringe capacitors.
\newblock \emph{IEEE Transactions on Circuits and Systems I: Regular Papers},
  61\penalty0 (8):\penalty0 2236--2242, 2014.

\bibitem[Eckert et~al.(2014)Eckert, Jayasena, and Loh]{thermal}
Yasuko Eckert, Nuwan Jayasena, and Gabriel~H Loh.
\newblock Thermal feasibility of die-stacked processing in memory.
\newblock 2014.

\bibitem[{Facebook AI Research}()]{caffe2}
{Facebook AI Research}.
\newblock Caffe2.
\newblock \url{https://caffe2.ai/}.

\bibitem[Krizhevsky(2014)]{alexnet}
Alex Krizhevsky.
\newblock One weird trick for parallelizing convolutional neural networks.
\newblock \emph{arXiv}, 2014.

\bibitem[Deng et~al.(2009)Deng, Dong, Socher, Li, Li, and Fei-Fei]{imagenet}
J.~Deng, W.~Dong, R.~Socher, L.-J. Li, K.~Li, and L.~Fei-Fei.
\newblock Imagenet: A large-scale hierarchical image database.
\newblock In \emph{CVPR}, 2009.
\newblock URL \url{http://image-net.org/}.

\bibitem[Simonyan and Zisserman(2014)]{vgg}
Karen Simonyan and Andrew Zisserman.
\newblock Very deep convolutional networks for large-scale image recognition.
\newblock \emph{arXiv}, 2014.

\bibitem[Hubara et~al.(2016)Hubara, Courbariaux, Soudry, El-Yaniv, and
  Bengio]{qnn:arxiv:2016}
Itay Hubara, Matthieu Courbariaux, Daniel Soudry, Ran El-Yaniv, and Yoshua
  Bengio.
\newblock Quantized neural networks: Training neural networks with low
  precision weights and activations.
\newblock \emph{arXiv}, 2016.

\bibitem[Krizhevsky and Hinton(2009)]{cifar10}
Alex Krizhevsky and Geoffrey Hinton.
\newblock Learning multiple layers of features from tiny images.
\newblock \emph{Computer Science Department, University of Toronto, Tech. Rep},
  2009.

\bibitem[Szegedy et~al.(2015)Szegedy, Liu, Jia, Sermanet, Reed, Anguelov,
  Erhan, Vanhoucke, and Rabinovich]{googlenet}
Christian Szegedy, Wei Liu, Yangqing Jia, Pierre Sermanet, Scott Reed, Dragomir
  Anguelov, Dumitru Erhan, Vincent Vanhoucke, and Andrew Rabinovich.
\newblock Going deeper with convolutions.
\newblock In \emph{Proceedings of the IEEE conference on computer vision and
  pattern recognition}, pages 1--9, 2015.

\bibitem[He et~al.(2016)He, Zhang, Ren, and Sun]{resnet}
Kaiming He, Xiangyu Zhang, Shaoqing Ren, and Jian Sun.
\newblock Deep residual learning for image recognition.
\newblock In \emph{CVPR}, 2016.

\bibitem[Redmon and Farhadi(2018)]{yolov3}
Joseph Redmon and Ali Farhadi.
\newblock Yolov3: An incremental improvement.
\newblock \emph{arXiv preprint arXiv:1804.02767}, 2018.

\bibitem[Marcus et~al.(1993)Marcus, Marcinkiewicz, and
  Santorini]{penn-treebank}
Mitchell~P Marcus, Mary~Ann Marcinkiewicz, and Beatrice Santorini.
\newblock Building a large annotated corpus of english: The penn treebank.
\newblock \emph{Computational linguistics}, 1993.

\bibitem[Hochreiter and Schmidhuber(1997)]{lstm}
Sepp Hochreiter and J{\"u}rgen Schmidhuber.
\newblock Long short-term memory.
\newblock \emph{Neural computation}, 1997.

\bibitem[Zhou et~al.(2016)Zhou, Ni, Zhou, Wen, Wu, and Zou]{dorefa:arxiv:2016}
Shuchang Zhou, Zekun Ni, Xinyu Zhou, He~Wen, Yuxin Wu, and Yuheng Zou.
\newblock Dorefa-net: Training low bitwidth convolutional neural networks with
  low bitwidth gradients.
\newblock \emph{arXiv}, 2016.

\bibitem[Mishra et~al.(2017)Mishra, Nurvitadhi, Cook, and Marr]{wrpn}
Asit~K. Mishra, Eriko Nurvitadhi, Jeffrey~J. Cook, and Debbie Marr.
\newblock {WRPN:} wide reduced-precision networks.
\newblock \emph{arXiv}, 2017.

\bibitem[Li et~al.(2016)Li, Zhang, and Liu]{li2016ternary}
Fengfu Li, Bo~Zhang, and Bin Liu.
\newblock Ternary weight networks.
\newblock \emph{arXiv}, 2016.

\bibitem[Zhang et~al.(2018)Zhang, Yang, Ye, and Hua]{zhang2018lq}
Dongqing Zhang, Jiaolong Yang, Dongqiangzi Ye, and Gang Hua.
\newblock Lq-nets: Learned quantization for highly accurate and compact deep
  neural networks.
\newblock \emph{arXiv preprint arXiv:1807.10029}, 2018.

\bibitem[ten()]{tensorrt}
Nvidia tensor rt 5.1.
\newblock \url{https://developer.nvidia.com/tensorrt}.

\bibitem[Song et~al.(2017)Song, Qian, Li, and Chen]{pipelayer:hpca:17}
Linghao Song, Xuehai Qian, Hai Li, and Yiran Chen.
\newblock Pipelayer: A pipelined reram-based accelerator for deep learning.
\newblock In \emph{High Performance Computer Architecture (HPCA), 2017 IEEE
  International Symposium on}, pages 541--552. IEEE, 2017.

\bibitem[NCSU(2018)]{freepdk}
NCSU.
\newblock Freepdk45, 2018.
\newblock URL \url{https://www.eda.ncsu.edu/wiki/FreePDK45}.

\bibitem[Murmann()]{survey}
B.~Murmann.
\newblock \emph{ADC Performance Survey 1997-2016}.
\newblock murmann/adcsurvey.html, [Online]. Available.
\newblock URL \url{http://web.stanford.edu/}.

\bibitem[Harpe(2018)]{cicc18:ADC}
Pieter Harpe.
\newblock A 0.0013 mm2 10b 10ms/s sar adc with a 0.0048 mm2 42db-rejection
  passive fir filter.
\newblock In \emph{2018 IEEE Custom Integrated Circuits Conference, CICC 2018}.
  Institute of Electrical and Electronics Engineers Inc., 2018.

\bibitem[Li et~al.(2011)Li, Chen, Ahn, Brockman, and Jouppi]{cactip}
S.~Li, K.~Chen, J.~H. Ahn, J.~B. Brockman, and N.~P. Jouppi.
\newblock {CACTI-P: Architecture-level Modeling for SRAM-based Structures with
  Advanced Leakage Reduction Techniques}.
\newblock In \emph{ICCAD}, 2011.

\bibitem[Paszke et~al.(2017)Paszke, Gross, Chintala, Chanan, Yang, DeVito, Lin,
  Desmaison, Antiga, and Lerer]{pytorch}
Adam Paszke, Sam Gross, Soumith Chintala, Gregory Chanan, Edward Yang, Zachary
  DeVito, Zeming Lin, Alban Desmaison, Luca Antiga, and Adam Lerer.
\newblock Automatic differentiation in pytorch.
\newblock In \emph{NIPS-W}, 2017.

\bibitem[Zmora et~al.(2018)Zmora, Jacob, and Novik]{distiller}
Neta Zmora, Guy Jacob, and Gal Novik.
\newblock Neural network distiller, June 2018.
\newblock URL \url{https://doi.org/10.5281/zenodo.1297430}.

\bibitem[Long et~al.(2018)Long, Na, and Mukhopadhyay]{long2018reram}
Yun Long, Taesik Na, and Saibal Mukhopadhyay.
\newblock Reram-based processing-in-memory architecture for recurrent neural
  network acceleration.
\newblock \emph{IEEE Transactions on Very Large Scale Integration (VLSI)
  Systems}, \penalty0 (99):\penalty0 1--14, 2018.

\bibitem[Crols and Steyaert(1994)]{scap:amp}
Jan Crols and Michel Steyaert.
\newblock Switched-opamp: An approach to realize full cmos switched-capacitor
  circuits at very low power supply voltages.
\newblock \emph{IEEE Journal of Solid-State Circuits}, 29\penalty0
  (8):\penalty0 936--942, 1994.

\bibitem[Fiorenza et~al.(2006)Fiorenza, Sepke, Holloway, Sodini, and
  Lee]{scap:converter}
John~K Fiorenza, Todd Sepke, Peter Holloway, Charles~G Sodini, and Hae-Seung
  Lee.
\newblock Comparator-based switched-capacitor circuits for scaled cmos
  technologies.
\newblock \emph{IEEE Journal of Solid-State Circuits}, 41\penalty0
  (12):\penalty0 2658--2668, 2006.

\bibitem[Brodersen et~al.(1979)Brodersen, Gray, and Hodges]{scap:filter}
Robert~W Brodersen, Paul~R Gray, and David~A Hodges.
\newblock Mos switched-capacitor filters.
\newblock \emph{Proceedings of the IEEE}, 67\penalty0 (1):\penalty0 61--75,
  1979.

\bibitem[Bankman and Murmann(2016)]{Mour:SCdotproduct}
Daniel Bankman and Boris Murmann.
\newblock An 8-bit, 16 input, 3.2 pj/op switched-capacitor dot product circuit
  in 28-nm fdsoi cmos.
\newblock In \emph{Solid-State Circuits Conference (A-SSCC), 2016 IEEE Asian},
  pages 21--24. IEEE, 2016.

\bibitem[Miyashita et~al.(2017)Miyashita, Kousai, Suzuki, and
  Deguchi]{timedomain:jssc2017}
Daisuke Miyashita, Shouhei Kousai, Tomoya Suzuki, and Jun Deguchi.
\newblock A neuromorphic chip optimized for deep learning and cmos technology
  with time-domain analog and digital mixed-signal processing.
\newblock \emph{IEEE Journal of Solid-State Circuits}, 52\penalty0
  (10):\penalty0 2679--2689, 2017.

\bibitem[Qiao et~al.(2018)Qiao, Cao, Yang, Song, and Li]{qiao2018atomlayer}
Ximing Qiao, Xiong Cao, Huanrui Yang, Linghao Song, and Hai Li.
\newblock Atomlayer: a universal reram-based cnn accelerator with atomic layer
  computation.
\newblock In \emph{Proceedings of the 55th Annual Design Automation
  Conference}, page 103. ACM, 2018.

\bibitem[Ji et~al.(2018)Ji, Song, Jiang, Li, and Chen]{ji2018recom}
Houxiang Ji, Linghao Song, Li~Jiang, Hai~Halen Li, and Yiran Chen.
\newblock Recom: An efficient resistive accelerator for compressed deep neural
  networks.
\newblock In \emph{Design, Automation \& Test in Europe Conference \&
  Exhibition (DATE), 2018}, pages 237--240. IEEE, 2018.

\bibitem[Li et~al.(2018)Li, Song, Chen, Qian, Chen, and Li]{li2018reram}
Bing Li, Linghao Song, Fan Chen, Xuehai Qian, Yiran Chen, and Hai~Helen Li.
\newblock Reram-based accelerator for deep learning.
\newblock In \emph{Design, Automation \& Test in Europe Conference \&
  Exhibition (DATE), 2018}, pages 815--820. IEEE, 2018.

\bibitem[Chen et~al.(2017{\natexlab{b}})Chen, Li, Chen, Deng, Shen, Liang, and
  Jiang]{chen2017accelerator}
Lerong Chen, Jiawen Li, Yiran Chen, Qiuping Deng, Jiyuan Shen, Xiaoyao Liang,
  and Li~Jiang.
\newblock Accelerator-friendly neural-network training: learning variations and
  defects in rram crossbar.
\newblock In \emph{Proceedings of the Conference on Design, Automation \& Test
  in Europe}, pages 19--24. European Design and Automation Association,
  2017{\natexlab{b}}.

\bibitem[Chi et~al.(2016{\natexlab{b}})Chi, Li, Xu, Zhang, Zhao, Liu, Wang, and
  Xie]{chi2016prime}
Ping Chi, Shuangchen Li, Cong Xu, Tao Zhang, Jishen Zhao, Yongpan Liu, Yu~Wang,
  and Yuan Xie.
\newblock Prime: A novel processing-in-memory architecture for neural network
  computation in reram-based main memory.
\newblock In \emph{ACM SIGARCH Computer Architecture News}, volume~44, pages
  27--39. IEEE Press, 2016{\natexlab{b}}.

\end{thebibliography}

\end{document}